\documentclass[twocolumn]{aastex7}

\usepackage{amsmath}
\usepackage{rotating}
\usepackage{comment}

\begin{document}

\title{Cloud-Cloud Collisions Induce Filament-Mediated Super Star Cluster Formation in the Antennae Overlap Region: Evidence from ALMA and JWST}

\author[orcid=0000-0003-2475-7983,gname=Tomonari, sname='Michiyama']{Tomonari Michiyama}
\affiliation{Faculty of Information Science, Shunan University, 843-4-2 Gakuendai, Shunan, Yamaguchi, 745-8566, Japan}
\email[show]{t.michiyama.astr@gmail.com}  

\author[orcid=0000-0002-2501-9328, gname=Toshiki, sname='Saito']{Toshiki Saito} 
\affiliation{Faculty of Global Interdisciplinary Science and Innovation, Shizuoka University, 836 Ohya, Suruga-ku, Shizuoka 422-8529, Japan}
\email{saito.toshiki@shizuoka.ac.jp}

\author[orcid=0000-0002-6939-0372, gname=Kouichiro, sname=Kouichiro]{Kouichiro Nakanishi}
\affiliation{National Astronomical Observatory of Japan, National Institutes of Natural Sciences, 2-21-1 Osawa, Mitaka, Tokyo, 181-8588, Japan}
\affiliation{Department of Astronomical Science, The Graduate University for Advanced Studies, SOKENDAI, 2-21-1 Osawa, Mitaka, Tokyo, 181-8588, Japan}
\email{nakanisi.k@nao.ac.jp}

\author[orcid=0000-0002-2364-0823, gname=Daisuke, sname=Iono]{Daisuke Iono}
\affiliation{National Astronomical Observatory of Japan, National Institutes of Natural Sciences, 2-21-1 Osawa, Mitaka, Tokyo, 181-8588, Japan}
\affiliation{Department of Astronomical Science, The Graduate University for Advanced Studies, SOKENDAI, 2-21-1 Osawa, Mitaka, Tokyo, 181-8588, Japan}
\email{d.iono@nao.ac.jp}

\author[orcid=0000-0002-2062-1600, gname=Kazuki, sname=Tokuda]{Kazuki Tokuda}
\affiliation{Faculty of Education, Kagawa University, Saiwai-cho 1-1, Takamatsu, Kagawa 760-8522, Japan}
\email{tokuda.kazuki@kagawa-u.ac.jp}

\author[orcid=0000-0002-2794-4840, gname=Kisetsu, sname=Tsuge]{Kisetsu Tsuge}
\affiliation{Faculty of Engineering, Gifu University, 1-1 Yanagido, Gifu, Gifu 501-1193, Japan}
\affiliation{Institute for Advanced Study, Gifu University, 1-1 Yanagido, Gifu, Gifu 501-1193, Japan }
\affiliation{Institute for Advanced Research, Nagoya University, Furo-cho, Chikusa-ku, Nagoya, Aichi 464-8601, Japan}
\email{tokuda.kazuki@kagawa-u.ac.jp}

\author[gname=Yuzuki, sname=Nagashima]{Yuzuki Nagashima}
\affiliation{Department of Astronomical Science, The Graduate University for Advanced Studies, SOKENDAI, 2-21-1 Osawa, Mitaka, Tokyo, 181-8588, Japan}
\affiliation{National Astronomical Observatory of Japan, National Institutes of Natural Sciences, 2-21-1 Osawa, Mitaka, Tokyo, 181-8588, Japan}
\email{yuzuki.nagashima@grad.nao.ac.jp}

\author[orcid=0009-0007-2493-0973, gname=Shinya, sname=Komugi]{Shinya Komugi}
\affiliation{Division of Liberal Arts, Kogakuin University, 2665-1 Nakano-cho, Hachioji, Tokyo 192-0015, Japan}
\email{skomugi@cc.kogakuin.ac.jp}

\begin{abstract}
The formation of super star clusters (SSCs) in galaxies remains a fundamental yet unresolved problem.
Among the proposed mechanisms, cloud–cloud collisions (CCCs) have been suggested as a potential trigger, although observational validation has been limited.
Here we present high-resolution ($0\farcs12$, $\sim$14~pc) ALMA observations of CO~($J=1$--$0$) emission toward a super giant molecular cloud (SGMC) in the overlap region of the Antennae galaxies. 
The data resolve the SGMC into two distinct velocity components separated by $\sim$50~km~s$^{-1}$. 
One component exhibits a ``U"-shaped structure within a large filament likely shaped by ram pressure, while the other shows hub–filament morphology.
Such a morphology is naturally interpreted as a CCC scenario.
The 108~GHz continuum emission detected at the apparent collision interface is dominated by free–free radiation, with an ionizing photon rate consistent with the stellar mass and age of the optically identified SSCs.
Supplementary infrared imaging with JWST reveals emission spatially coincident with the inferred collision interface, further supporting the CCC scenario.
These results provide compelling, multi-wavelength evidence that CCCs play a key role in triggering SSC formation in merging galaxies.
\end{abstract}

\keywords{
\uat{Extragalactic astronomy}{506} --- 
\uat{Galaxy collisions}{585} --- 
\uat{Galaxy interactions}{600} --
\uat{Galaxy mergers}{608} --
\uat{Interstellar clouds}{834} --
\uat{Millimeter astronomy}{1061} --
\uat{Molecular clouds}{1072} --
\uat{Star forming regions}{1565} --
\uat{Starburst galaxies}{1570} --
\uat{Young star clusters}{1833} --
}


\section{Introduction} \label{sec:intr}
Among nearby merging galaxies, the Antennae galaxies (NGC~4038/4039) provide one of the clearest examples of ongoing star formation \citep[e.g.,][]{Renaud2014, Renaud2019}.  
High-resolution Hubble Space Telescope (HST) imaging has revealed that much of the ongoing star formation takes place in massive, compact clusters \citep[e.g.,][]{Whitmore1995, Whitmore1999}.  
The so-called ``overlap" region, where the two galactic disks interpenetrate, hosts intense star formation and a rich population of young ($<10$~Myr) super star clusters (SSCs) embedded in a turbulent, dust-rich environment \citep{Mengel2005, Whitmore2010}.
Since star formation is fundamentally governed by the properties of molecular gas, characterizing its spatial distribution, physical conditions, and kinematics is essential for understanding how SSCs form in merging systems.

Interferometric CO observations have revealed massive molecular structures known as super giant molecular clouds (SGMCs) \citep{Wilson2000}.  
Using $\sim$1$^{\prime\prime}$-resolution CO~($J=3$--$2$) data from the Submillimeter Array, \citet{Ueda2012} identified 57 molecular complexes, more than half of which are located in the overlap region.  
Subsequent ALMA observations, with their superior resolution and sensitivity, have uncovered intricate substructures and faint extended emission, demonstrating that the molecular gas is under extremely high pressure ($P/k_{B} \gtrsim 10^8$~K~cm$^{-3}$) conditions conducive to the formation of massive clusters \citep{Johnson2015, Finn2019}.

Although the physical mechanism that triggers the formation of massive clusters in merging systems remains unclear,
the so-called overlap region, by definition, represents the interface where kiloparsec-scale shocks occur,
and thus naturally provides a dense and highly pressurized environment conducive to cluster formation \citep{Jog1992}.
Numerical simulations and multi-scale observations have shown that collisions between molecular clouds can serve as an efficient trigger of star formation \citep[e.g.,][]{Tan2000, Tasker2009, Inoue2013}.  
Such interactions enhance the gas density and external pressure, favoring the formation of massive clusters \citep{Elmegreen1997, Fukui2014}.
Although tracing the detailed trajectories and collision histories of individual clouds remains challenging in galactic-scale simulations, the presence of highly clumpy and filamentary molecular gas in the overlap region provides indications that cloud–cloud interactions could be occurring in the Antennae galaxies \citep{Renaud2019}.

Motivated by this theoretical expectation, several observational studies have sought signatures of such collisions in the system.
In particular, \citet{Tsuge2021, Tsuge2021_B1} presented compelling evidence for SSC formation triggered by cloud–cloud collisions (CCCs), based on ALMA CO~($J=3$--$2$) data.  
Their work provided a systematic view of CCC signatures across the overlap region, yet the available resolution and sensitivity were insufficient to fully resolve the internal structure and dynamics of the interacting clouds.

In this study, we report on a remarkable SGMC in the Antennae galaxies, where multiple lines of evidence converge: an optically identified SSC, strong free–free continuum emission detected with ALMA, and two well-separated molecular gas components exhibiting signatures of a CCC. This system offers a unique opportunity to investigate the interplay between molecular gas dynamics, ongoing star formation, and the birth of massive stellar clusters.

We adopt a fiducial redshift of $z = 0.005474$ for the Antennae galaxies.  
The corresponding distance, corrected for infall toward the Virgo Cluster, the Great Attractor, and the Shapley Supercluster as reported by the NASA/IPAC Extragalactic Database, is taken to be $D = 25$~Mpc.  
At this distance, 1$^{\prime\prime}$ corresponds to 122~pc.

\section{Data}
\begin{sidewaysfigure}
\centering
\includegraphics[width=0.99\textwidth]{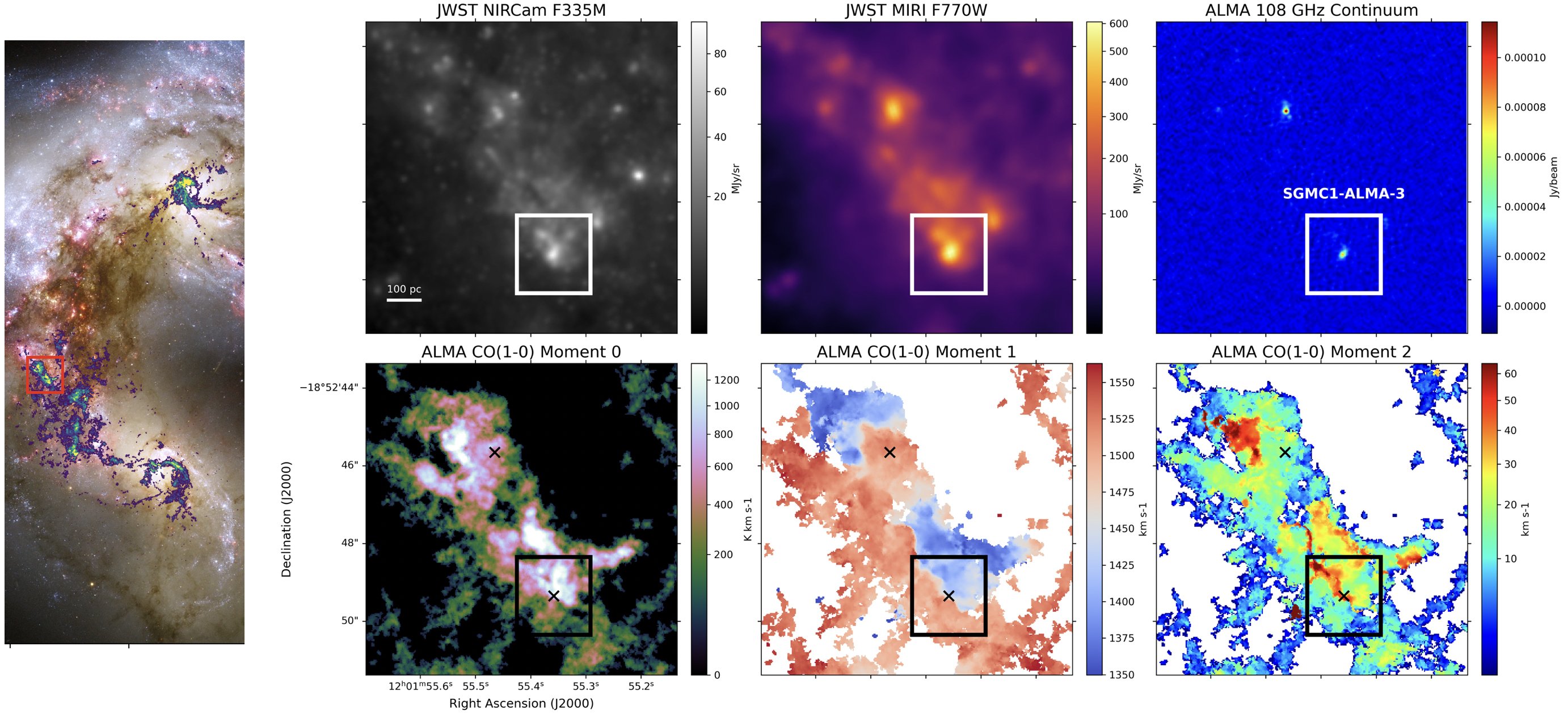}
\caption{
The leftmost panel shows an HST color composite image with ALMA CO~($J=1$--$0$) moment~0 maps, with the red box indicating the SGMC1 region. 
The six panels to the right provide a zoom-in view of this region. The top row displays the JWST NIRCam F335M image, the JWST MIRI F770W image, and the ALMA 108~GHz continuum map, arranged from left to right. 
The bottom row shows the corresponding ALMA CO~($J=1$--$0$) maps: moment~0 (integrated intensity), moment~1 (intensity-weighted velocity), and moment~2 (velocity dispersion), also arranged from left to right. 
The white and black squares in the top and bottom panels, respectively, indicate the same $2\farcs \times 2\farcs$ region centered on the secondary peak of the ALMA continuum emission (R.A. $= 180.480659$~degree, Decl. $= -18.880374$~degree), corresponding to SGMC1-ALMA-3, which is the focus of this study. Black crosses mark the positions of the ALMA continuum peaks. A 100~pc scale bar is shown in the JWST F335M panel.
\label{fig:panel}}
\end{sidewaysfigure}

Figure~\ref{fig:panel} shows ALMA and James Webb Space Telescope (JWST) images obtained from the archive. 
In this paper, we focus on the SGMC1 region, defined as an 8 arcsec box centered at (R.A., Decl.) = (180.4808962~degree, $-18.8798212$~degree), as identified by \citet{Wilson2000}. 
We also use the catalog of SSCs compiled by \citet{Whitmore2010}. 
Within SGMC1, \citet{Whitmore2014} identified a radio emission peak with SSC candidates labeled as ``SGMC1-ALMA-3", which is the main focus of this study.

\subsection{ALMA}
This study makes use of new ALMA data from the Observatory Project 2022.A.00032.S.  
The program was designed to map the molecular gas distribution across the entire Antennae system at high spatial resolution, with the goal of characterizing molecular cloud properties in various dynamical environments.  
The CO~($J=1$--$0$) mosaic observations cover both galactic disks and the overlap region where the two disks interact (the leftmost panel of Figure~\ref{fig:panel}), using the 12-m array.  
This wide mosaic encompasses the overlap region containing several SGMCs in literature \citep{Wilson2000}.  
In the present study, we focus on a small subregion around SGMC1, which exhibits active star formation and hosts several young super star clusters \citep{He22}.  

The observations were conducted between September and November 2023, as part of ALMA Cycle 9 and 10.
Details of the data calibration and imaging procedures are described in a separate paper (Saito et al., in preparation).  
The resulting CO~($J=1$--$0$) data cube achieves an angular resolution of $0\farcs12$ (14.1~pc) and a velocity resolution of 2.54~km~s$^{-1}$, with a typical noise level of 2.1~K ($1\sigma$) measured in line-free channels within the SGMC1 region.  
This dataset has not been analyzed in any previous publication.

\subsection{JWST}
We downloaded the JWST data\footnote{The file names are \texttt{jw02581-o001\_t001\_nircam\_clear-f335m} and \texttt{jw02581-o002\_t002\_miri\_f770w}.  All the {\it JWST}, data used in this paper can be found in MAST: \dataset[10.17909/vm9w-vk88]{http://dx.doi.org/10.17909/vm9w-vk88}.} (NIRCam and MIRI imaging) from the archive under Proposal ID 2581.
Since detailed flux measurements based on the JWST data are beyond the scope of this study, we focus only on the morphology in the NIRCam F335M and MIRI F770W filters, both of which are sensitive to polycyclic aromatic hydrocarbon (PAH) emission \citep{Sutter2024}.

\subsection{SSC Catalog}
To investigate the spatial distribution of SSCs in SGMC1, we use Tables 6, 7, and 8 from \citet{Whitmore2010}, which list the 50 most luminous, most massive, and most IR-bright clusters, respectively.
We identified four SSC candidates within SGMC1-ALMA-3  (Table~\ref{tab:SSCs}).
This region corresponds to the position ``D" in \citet{Tsuge2021}.

\begin{deluxetable*}{lcccc}
\tablewidth{0pt}
\tablecaption{SSCs in SGMC1-ALMA-3 \label{tab:SSCs}}
\tablehead{
\colhead{ID} & \colhead{$\log_{10}\tau_{\rm cl}$} & \colhead{$M_{\rm cl}$} & \colhead{$E(B-V)$} & \colhead{Table} \\
\colhead{} & \colhead{yr} & \colhead{$M_\odot$} & \colhead{} & \colhead{} 
}
\colnumbers
\startdata
19275\tablenotemark{a} & 6.02 & $4.68 \times 10^5$ & 0.320 & 6 \\
19459 & 6.56 & $1.25 \times 10^5$ & 0.100 & 6 \\
19416 & 6.72 & $8.76 \times 10^4$ & 0.000 & 6 \\
19330\tablenotemark{a} & 6.16 & $3.20 \times 10^5$ & 0.340 & 8 \\
\enddata
\tablenotetext{a}{SSCs located within the ALMA 108~GHz continuum emission.}
\tablecomments{
(1) IDs from \citet{Whitmore2010}; 
(2) logarithmic cluster ages in years; 
(3) cluster masses in solar masses; 
(4) color excess $E(B-V)$ indicating reddening; 
(5) corresponding table number in \citet{Whitmore2010}.
}
\end{deluxetable*}

\section{Results}
In the following sections, we present a detailed investigation of the SGMC1-ALMA-3 region.
As shown in Figure~\ref{fig:panel}, this position—characterized by a large velocity dispersion—is located at the interface between two molecular filaments: one blue-shifted ($\lesssim1400$~km\,s$^{-1}$) and one red-shifted ($\gtrsim1500$~km\,s$^{-1}$).
This region hosts optically identified SSCs, exhibits strong free–free continuum emission detected by ALMA, and contains abundant molecular gas, making it a valuable snapshot of massive cluster formation.
The observed gas morphology and kinematics around SGMC1-ALMA-3 further show several features that are consistent with theoretical predictions of CCCs \citep{Takahira2014ApJ...792...63T, Haworth2015MNRAS.454.1634H, Haworth2015MNRAS.450...10H}, including a ``U"-shaped structure, hub–filament morphology, and a bridge-like feature in the position–velocity (PV) diagram, which we discuss in more detail in Section~\ref{sec:ccc}.
These properties make SGMC1-ALMA-3 an ideal target for investigating potential signatures of CCCs, and we focus our analysis on this region throughout the paper. 

\subsection{CO~($J=1$--$0$) cube}
\begin{figure*}[ht!]
\centering
\includegraphics[scale=0.35]{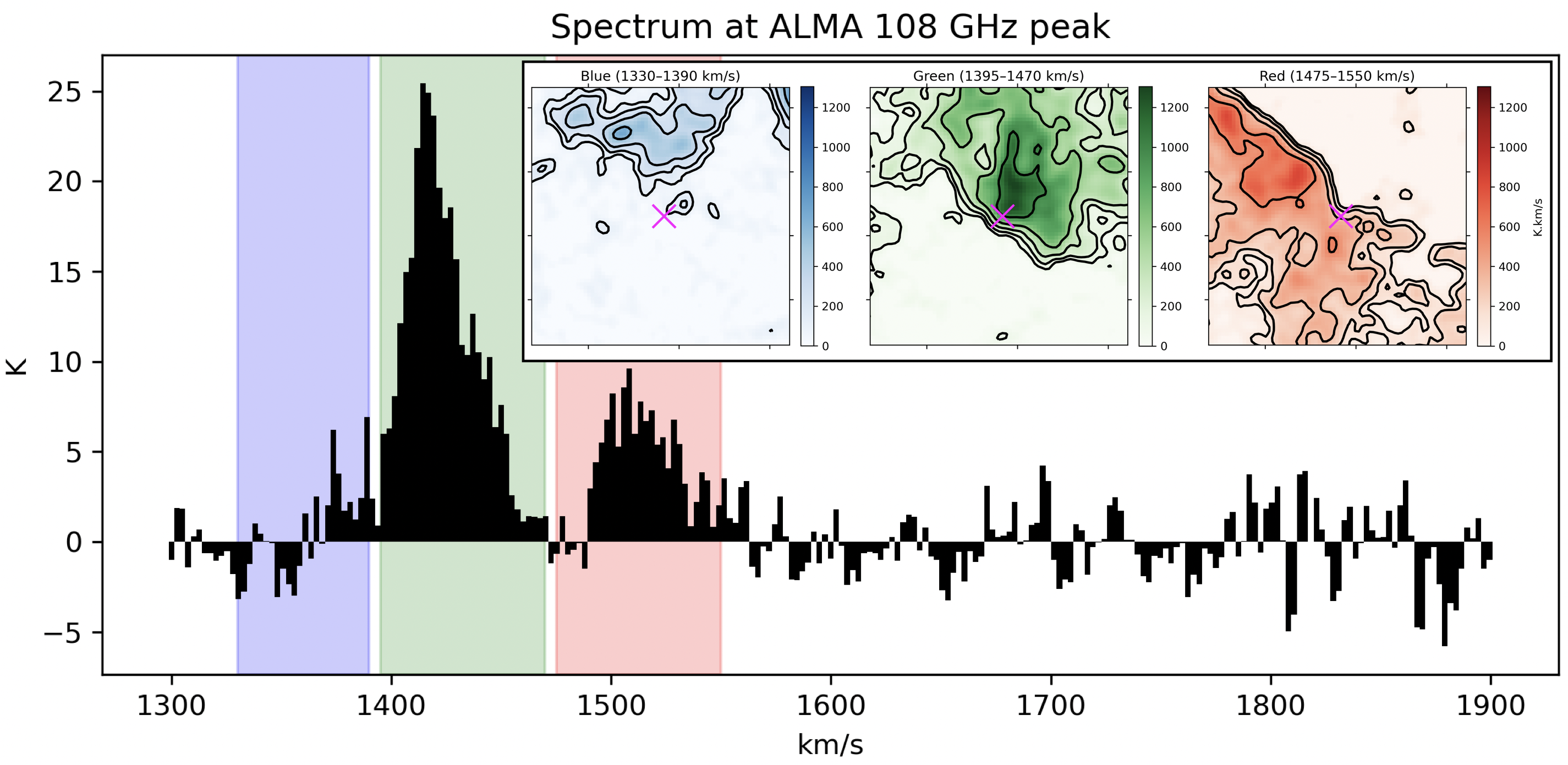} 
\caption{
(outer panel) CO~($J=1$--$0$) spectrum extracted at the ALMA 108~GHz peak pixel of SGMC1-ALMA-3. 
The background shading highlights three distinct velocity intervals: blue (1330--1390\,km\,s$^{-1}$), green (1395--1470\,km\,s$^{-1}$), and red (1475--1550\,km\,s$^{-1}$).
(inner panels) CO~($J=1$--$0$) integrated intensity maps of SGMC1-ALMA-3, shown for three velocity components: 1330–1390\,km\,s$^{-1}$ (blue, left), 1395–1470\,km\,s$^{-1}$ (green, center), and 1475–1550\,km\,s$^{-1}$ (red, right). The contour levels are set at 5 logarithmically spaced intervals between 100 and 1000~K~km~s$^{-1}$ to highlight both faint and bright emission features.
\label{fig:spec}}
\end{figure*}

\begin{figure*}[ht!]
\centering
\includegraphics[scale=0.46]{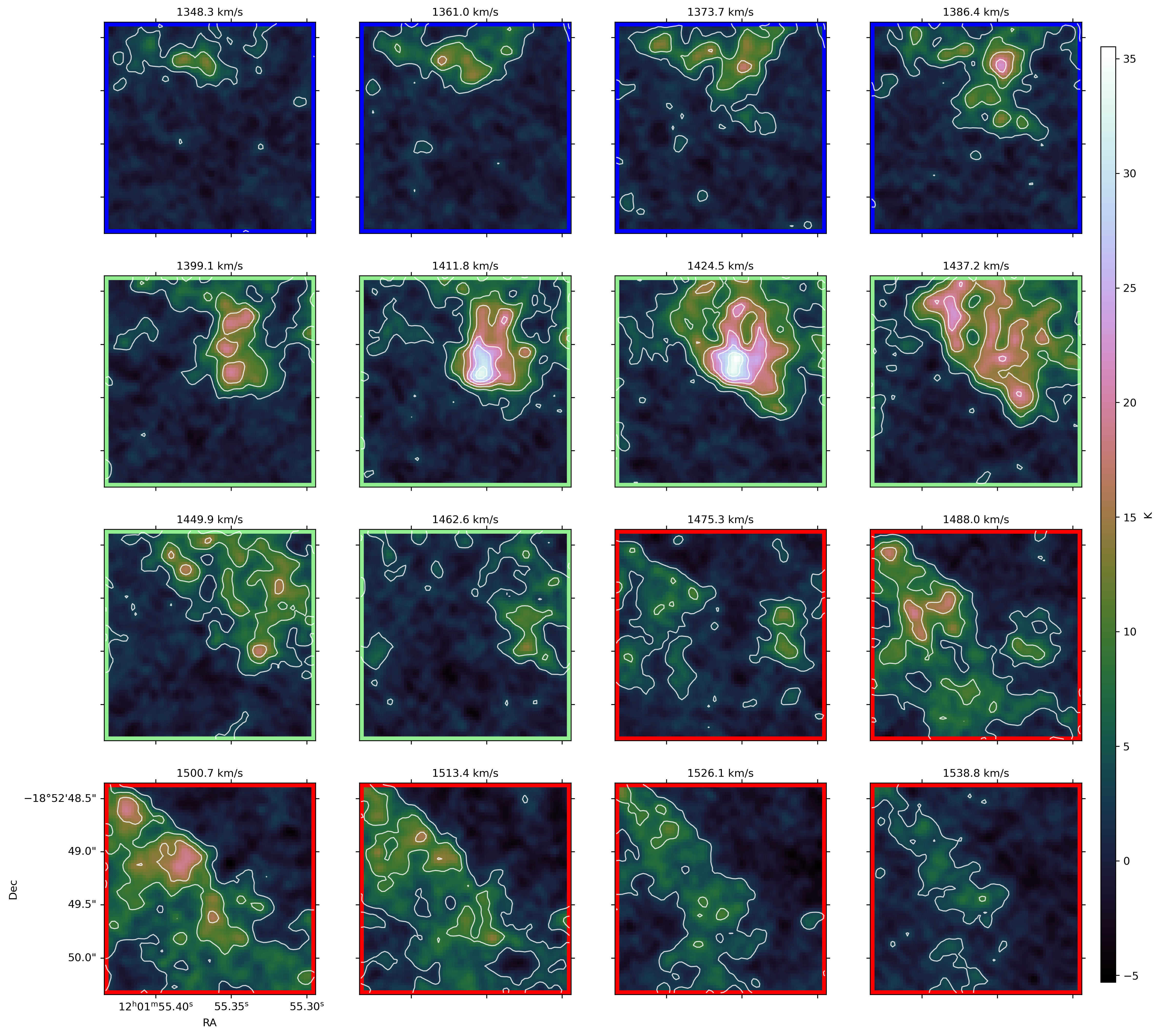} 
\caption{
{Channel-map sequence of the CO (1--0) emission cube toward SGMC1--ALMA-3.
Each panel shows a $2\farcs\times2\farcs$ field of view, derived from the cube binned by five channels along the spectral axis, with the velocity (km~s$^{-1}$) indicated at the top.
Contours indicate emission at 3, 9, 15, 21, 27, and 33 times the rms noise level,
where the rms noise per channel is estimated to be $2.1\,{\rm K}/\sqrt{5}$ after five-channel binning.
Colored borders delineate three characteristic velocity intervals (blue, green, and red) corresponding to the distinct gas components identified in Figure~\ref{fig:spec}.}
\label{fig:cmap}}
\end{figure*}

Figure~\ref{fig:spec} shows the CO~($J=1$--$0$) spectrum extracted at the position of SGMC1-ALMA-3.  
A multi-component spectrum is clearly observed.  
We define three velocity ranges to characterize the components.  
The main emission is concentrated in the green range (1395--1470\,km\,s$^{-1}$), while an additional redshifted component is evident in the red range (1475--1550\,km\,s$^{-1}$).  
Although blueshifted gas is not significantly detected at the exact location of SGMC1-ALMA-3,  Figure~\ref{fig:cmap} shows that such gas is distributed in the surrounding region.  
We therefore define the blue range as 1330--1390\,km\,s$^{-1}$.

\subsection{Composite View of CO~($J=1$--$0$) Emission, 108~GHz Continuum, and SSC Candidates}
\begin{figure*}[ht!]
\centering
\includegraphics[scale=0.30]{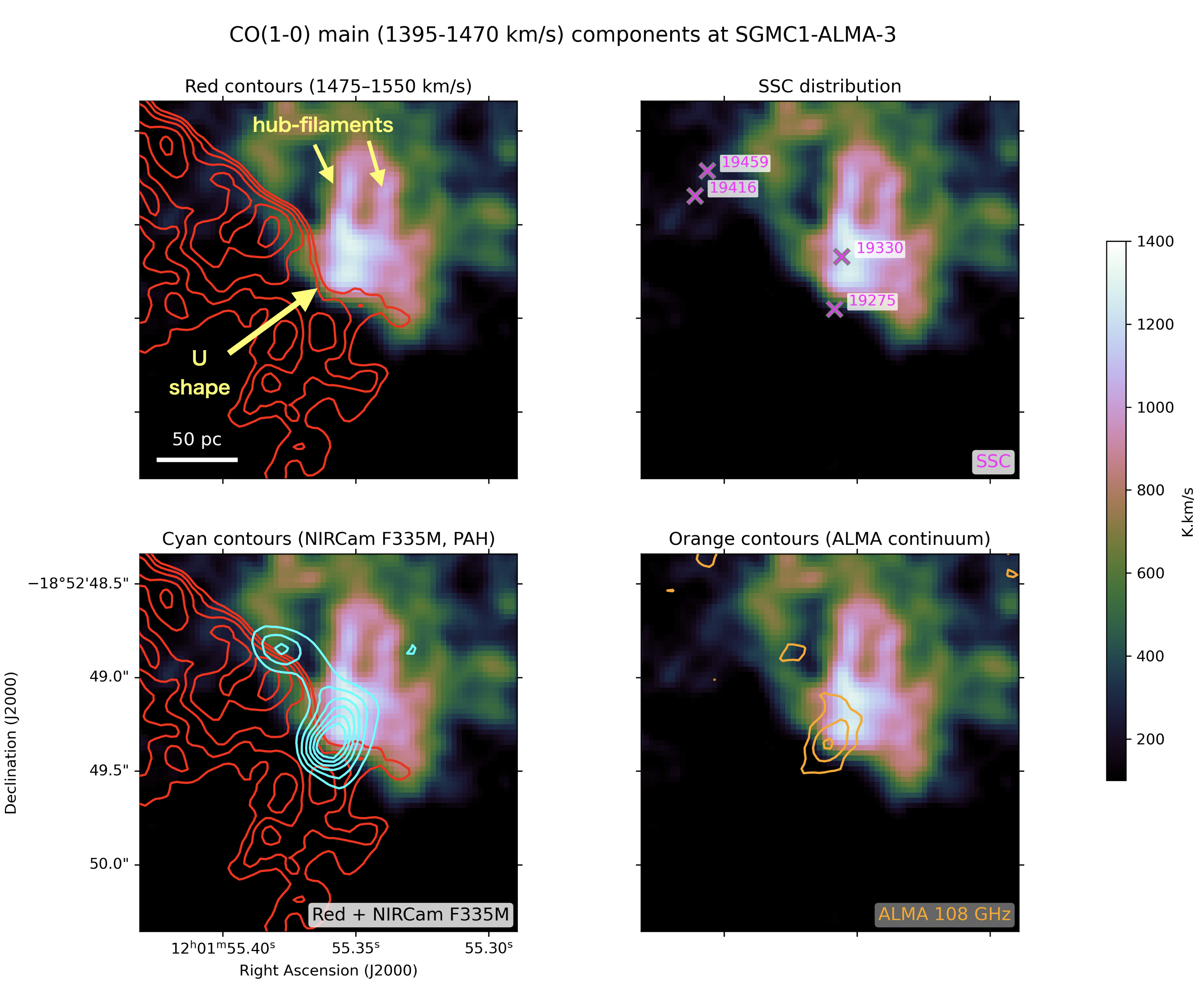} 
\caption{
Zoomed-in view of the SGMC1–ALMA-3 region, showing the CO~($J=1$--$0$) integrated intensity (moment 0) map at 1395–1470 km s$^{-1}$ with various overlays.
Top-left: Red contours trace the CO~($J=1$--$0$) emission integrated over 1475–1550 km s$^{-1}$ (contour levels: six logarithmically spaced levels between 300 and 1000 K km s$^{-1}$).
Top-right: Locations of optically identified SSC candidates from \citet{Whitmore2010}, marked with ‘×’ symbols and annotated with ID.
Bottom-left: Cyan contours indicate JWST/NIRCam F335M emission (contour levels: 0.5–1.2 times the 99.9th percentile of the image in 6 steps), overlaid with the same red CO contours as in the top-left panel.
Bottom-right: Orange contours show the ALMA 108 GHz continuum emission (contour levels: five logarithmically spaced levels between $1\times10^{-5}$ and $5\times10^{-4}$~Jy~beam$^{-1}$).
\label{fig:green}}
\end{figure*}

Figure~\ref{fig:green} presents four panels of the CO~($J=1$--$0$) integrated intensity map corresponding to the green velocity component. The bottom-right panel shows that the ALMA 108~GHz continuum peak lies at the interface between the red filament and the main (green) component. JWST/NIRCam F335M emission is also detected along the northeastern edge of the red-shifted filament.

In addition to the young SSC candidates within the 108~GHz peak --- 
ID = 19275 and 19330
--- the figure also shows relatively older SSCs:  
ID = 19459 and ID = 19416   
which are located farther to the northeast, beyond the northeastern edge of the NIRCam F335M emission.

\section{Discussion}
Following the classification scheme proposed by \citet{Whitmore2014}, star cluster formation proceeds through six stages: Stage~0 corresponds to diffuse giant molecular clouds; Stage~1 is the protocluster phase, younger than 0.1~Myr; Stage~2 is the embedded cluster phase, spanning 0.1--1~Myr; Stage~3 represents emerging clusters aged between 1 and 3~Myr; Stage~4 includes young clusters between 3 and 10~Myr; and Stage~5 corresponds to intermediate or old clusters older than 10~Myr. According to this scheme, SGMC1-ALMA-3 is classified as Stage~3 in \citet{Whitmore2014}. However, their study does not explore the detailed gas dynamics or the triggering mechanisms of massive star cluster formation at this stage.
In the following subsection, we discuss how CCCs contribute to the formation of SSCs at this stage.

\subsection{Cloud--Cloud Collision (CCC)}\label{sec:ccc}
We initially aimed to understand the physical mechanism of star formation traced by the ALMA 108~GHz continuum.
Inspection of the CO (1–0) channel maps revealed a remarkably complex velocity field.
As shown in Figure~\ref{fig:cmap}, the large velocity spread cannot be attributed to a simple explosive motion (e.g., a star formation–driven outflow).
Instead, multiple velocity components are spatially overlapping along the line of sight: red, green, and blue, indicating that distinct molecular structures are interacting within a confined region.
Such spatial and kinematic complexity is characteristic of regions where molecular clouds are colliding.
Motivated by these findings, we assess whether the observed structures are consistent with the expectations from the CCC scenario.
While the general concept of CCC has been introduced in Section~\ref{sec:intr}, in the following discussion we use the term to specifically refer to a localized interaction between two discrete molecular clouds on spatial scales of $\sim$50--100~pc, with the actual compressed interface typically being smaller (a few to tens of parsecs).

Indeed, numerical simulations have shown that CCCs produce a set of characteristic morphological and kinematic signatures: two distinct velocity components connected by a low-intensity “bridge” feature in position–velocity (PV) space, complementary spatial distributions of the colliding clouds, and localized enhancements of velocity dispersion or thermal pressure at the interface \citep[see the review by][]{Fukui2021}.
Similar signatures have been widely identified in Galactic regions: for instance, a U-shaped cavity in RCW~120 \citep{Torii2015} and a bridge feature in M~20 \citep{Torii2017}. These features are interpreted as evidence of dynamic gas compression and massive star formation triggered by CCCs.
In the following, we examine each of the observational signatures of CCCs in SGMC1-ALMA-3.

\subsubsection{``U"-shaped structure}
A ``U''-shaped feature is observed in the redshifted filament at the position of the ALMA 108~GHz continuum peak (Figure~\ref{fig:green}), which may be interpreted as compression caused by ram pressure from interaction with the main (green) component. When the external ram pressure, $P_{\rm ram} = \rho v^2$, exceeds the internal turbulent and thermal pressures, the filament can be deformed.

Using an aperture with a radius of
$R = 12.5~\mathrm{pc}$ we measure a mass of $M \approx 10^6~M_\odot$ (derived with a standard CO-to-H$_2$ conversion factor of 
$4.3~M_{\odot}\,\mathrm{(K\,km\,s^{-1}\,pc^{2})^{-1}}$; \citealt{Bolatto2013})
\footnote{The aperture is centered at a position shifted by one radius along $PA = 50^{\circ}$, such that the circle passes through the ALMA continuum peak.}.
If this mass is uniformly distributed within a sphere of radius $R$, the corresponding average volume density is
\begin{equation}
\rho = \frac{M}{\tfrac{4}{3}\pi R^3} \approx 8.3 \times 10^{-21}~\mathrm{g~cm^{-3}}.
\end{equation}
Adopting a relative velocity of $v = 50~\mathrm{km~s^{-1}}$, the ram pressure is estimated as $P_{\rm ram} = \rho v^2 \approx 1.7 \times 10^{-7}~\mathrm{dyne~cm^{-2}}$, which corresponds to a thermal pressure of $P_{\rm ram}/k_{\rm B} \approx 1.5 \times 10^9~\mathrm{K~cm^{-3}}$.

This value exceeds typical internal pressures in molecular clouds ($\sim10^4$--$10^6~\mathrm{K~cm^{-3}}$; e.g., \citealt{Schinnerer2024ARA&A..62..369S}), indicating that the incoming clump can dynamically compress the filament. Such compression is likely to enhance the local gas density and induce gravitational collapse, potentially triggering the observed star formation at the interface.

According to the empirical relation between external pressure and the total stellar mass of SSCs presented in \citet{Tsuge2021}, such a high pressure of $P_{\rm ram}/k_{\rm B} \approx 1.5 \times 10^9~\mathrm{K~cm^{-3}}$ is capable of forming a stellar cluster with mass $\sim 10^7~M_\odot$. However, the observed stellar mass of SSC $\lesssim 10^6~M_\odot$, suggesting that this region is still in an early phase of the collision. It is likely that a more massive structure will form as the surrounding gas becomes more involved in the interaction, potentially giving rise to a more massive SSC as the collision progresses.


\subsubsection{PV diagram}
\begin{figure*}[ht!]
\centering
\includegraphics[scale=0.18]{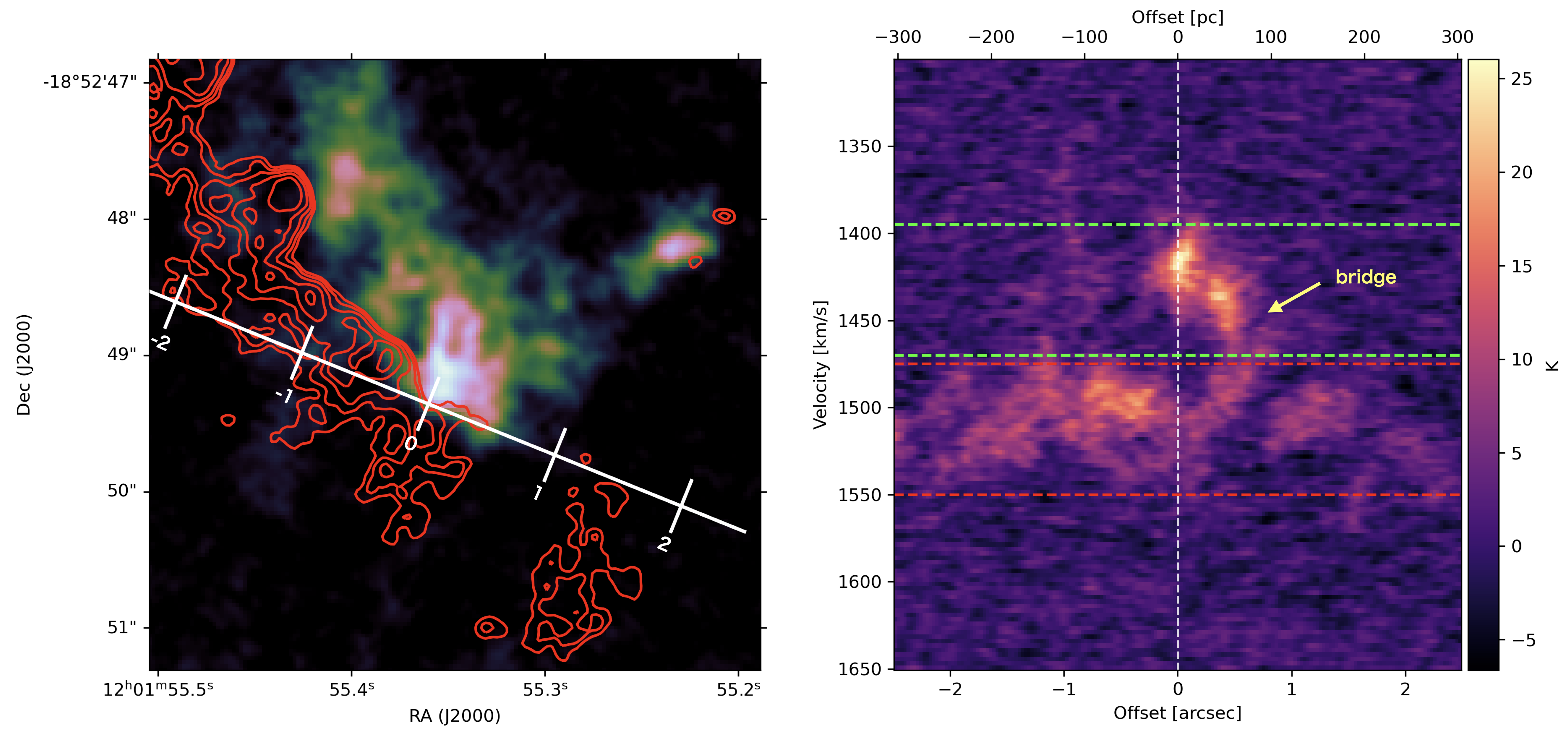} 
\caption{
Left: Same background image and red contours as in Figure~\ref{fig:green}, but with a wider field of view. The white line indicates the cut used to extract the PV diagram, and tick marks along the line denote offset positions at $-2$, $-1$, $0$, $+1$, and $+2$ arcsec.
Right: PV diagram extracted along the white line shown in the left panel, using a slit width of 1~pixel.
The horizontal dashed green and red lines indicate the velocity range, corresponding to Figure~\ref{fig:spec}.
\label{fig:pv}}
\end{figure*}

The PV diagram extracted along the white line shown in Figure~\ref{fig:pv} displays a bridge-like feature connecting two velocity components (the red-shifted filament and the main green component), which is a characteristic signature of cloud–cloud collisions. This velocity separation, together with the connecting low-intensity emission, is consistent with features predicted by numerical simulations of CCCs \citep[e.g.,][]{Haworth2015MNRAS.450...10H}.

These signatures support a scenario in which an ongoing collision between two molecular components creates a compressed layer of gas where massive stars and the SSC are forming. The observed PV structure thus provides kinematic evidence that reinforces the CCC interpretation initially suggested by the morphological features.

\subsubsection{Hub–filaments}
The main (green) component exhibits a hub–filament structure (Figure~\ref{fig:green}), with a characteristic filament width of approximately 14~pc (comparable to the beam size). 
Such a configuration is shown to arise in CCC scenarios in numerical simulations \citep{Inoue2018PASJ...70S..53I,Maity2024ApJ...974..229M}, and observationally, systems exhibiting active star formation at the tips of hub–filament structures have been found in compact environments such as the Large Magellanic Cloud \citep[e.g.,][]{Fukui2019ApJ...886...14F,Tokuda2019ApJ...886...15T,Sewi_2023ApJ...959...22S}.
In contrast, our observations reveal a similar hub–filament configuration associated with a massive ($\sim 10^6~M_\odot$) molecular structure, suggesting that such collision-induced morphology can also emerge on cluster-forming scales.

In this framework, the hub draws in large quantities of gas along its attached filaments during the earliest evolutionary phase, amassing the material reservoir that initiates stellar‑cluster formation and ultimately enables the birth of massive stars.
The observed morphology, in conjunction with the dynamical signatures of collision and the presence of active massive star formation, is consistent with this picture. 

\subsection{From Merger-Driven Flows to Cloud–Cloud Collision}
On kiloparsec scales, the molecular gas is dominated by the ongoing galaxy--galaxy interaction, which drives strong tidal forces, large-scale streaming motions, and enhanced turbulence.
The CCC scenario discussed in Section \ref{sec:ccc} does not exclude the influence of large-scale, merger-driven gas flows.
Rather, we consider that the ongoing galaxy--galaxy interaction in the Antennae may have created the dynamical conditions necessary for close encounters between individual molecular clouds within the SGMC1--ALMA--3 complex.

The key question is whether the observed small-scale structures in SGMC1--ALMA--3 are more naturally explained by a local interaction between two molecular clouds, or by random turbulent or streaming motions associated with the merging system.
Several lines of evidence favor the former interpretation.

First, within a region of $\sim$100~pc, we identify two CO components separated by $\Delta v \simeq 50$--100~km~s$^{-1}$, significantly larger than the linewidth expected for a single turbulent GMC of comparable size \citep{Solomon1987}, as well as those predicted by theoretical models of starburst mergers \citep{Bournaud2015}.
A comparative analysis by \citet{Brunetti2021,Brunetti2022,Brunetti2024} at $\sim$100~pc resolution indicates that the velocity dispersion in the more advanced merger NGC~3256 may be dominated by merger-driven flows, whereas similar dispersions in the Antennae are less pronounced.
The extreme dispersions observed near the nuclei of NGC~3256 are also likely associated with molecular outflows \citep{Sakamoto2014, Michiyama2018}.
In contrast, at the higher spatial resolution ($\sim10$~pc) achieved in our observations, the observed channel-map morphology and large velocity separations cannot be straightforwardly explained by merger-driven flows alone.

Second, the projected distributions of the two components exhibit a complementary pattern, in which the intensity peak of one component coincides with a depression in the other, consistent with partial overlap between finite-sized clouds.

Third, the two components are connected by a low-intensity ``bridge'' feature in position–velocity space, and the CO linewidth is significantly enhanced at their interface, coincident with the location of the young SSC.

The observed combination of a low-intensity bridge feature in position--velocity space and a localized enhancement of the CO linewidth at the interface is consistent with the behavior seen in numerical simulations of CCCs.
In contrast, turbulent motions or large-scale shear typically lead to broader and more spatially extended velocity structures.

Overall, the observed morphological and kinematic properties of SGMC1--ALMA--3 are consistent with a scenario in which large-scale merger-driven gas dynamics have created the conditions for a localized cloud--cloud interaction, possibly triggering the formation of the observed SSC.
Although this interpretation is not conclusive, it provides a self-consistent framework for explaining both the spatial distribution and the velocity structure of the molecular gas.
Similar connections between galaxy-scale tidal flows and localized gas interactions have also been discussed in other interacting systems, such as the Magellanic system, where tidally driven gas flows may contribute to the formation of compressed regions and subsequent star formation \citep{Fukui2017PASJ, Tokuda2022, Tsuge2024PASJ}.

\subsection{Free-free emission}
As discussed above, if ongoing star formation is being triggered by a CCC, the ALMA 108~GHz continuum emission observed at the interface between the red and green velocity components is expected to originate from free–free radiation produced by young massive stars.
This continuum source corresponds to  ``Index 5" in the catalog of embedded clusters presented by \citet{He22}.
In the following, we present observational evidence supporting this interpretation.
\subsubsection{Origin of 108~GHz continuum emission}
We performed 2D Gaussian fitting to the continuum source in SGMC1-ALMA-3 using the {\tt imfit} task within CASA. The deconvolved source size is $0\farcs22 \times 0\farcs11$ with a position angle of $150^\circ$. The integrated flux density derived from the fit is $245 \pm 20~\mu$Jy, and the peak intensity is $95 \pm 6~\mu$Jy beam$^{-1}$.
For comparison, the {\tt imstat} task applied to the same elliptical region (with a semi-major and semi-minor axis of $0\farcs35$) yields a total flux density of $308~\mu$Jy and a peak value of $83~\mu$Jy beam$^{-1}$.
Since the fitted flux can vary with imaging parameters (e.g., uv-taper, weighting scheme) and the fitting method, we adopt a representative flux density of $S_\nu \approx300~\mu$Jy at 100~GHz for the following discussion. As the source is not perfectly Gaussian, such a fit may introduce a small additional uncertainty.

The monochromatic luminosity \( L_\nu \) is computed as:
\begin{equation}
L_\nu = 4\pi D^2 S_\nu 
= 1.73 \times 10^{26}~\mathrm{erg\,s^{-1}\,Hz^{-1}}
\end{equation}
While the ALMA Band~6 and 7 continuum detections ($\sim800~\mu$Jy at 258GHz and $\sim1000~\mu$Jy at 337~GHz) may indicate thermal dust emission, the observed 108~GHz flux cannot be explained by a typical dust spectral energy distribution \citep[see also][]{He22}.
Furthermore, the radio map presented in \citet{Whitmore2014}, which has a larger beam size than ALMA Band~3, shows $\sim500~\mu$Jy\,beam$^{-1}$ at 8.6~GHz near the 108~GHz peak position.
If the 108~GHz emission were attributed to synchrotron radiation, it would require an unusually flat spectral index of $>-0.3$.
We therefore attribute the 100GHz continuum to free–free emission from ionized gas associated with massive star formation.

\subsubsection{The ionizing photon rate}
The ionizing photon rate $Q(\mathrm{H}^0)$ can be estimated from the thermal free–free continuum luminosity using the formulation of \citet{Rubin1968} and \citet{Murphy2011}:
\begin{multline}
Q(\mathrm{H}^0) = 6.3 \times 10^{25}
\left( \frac{T_e}{10^4\,\mathrm{K}} \right)^{-0.45}
\left( \frac{\nu}{\mathrm{GHz}} \right)^{0.1} \\
\times \left( \frac{L_\nu}{\mathrm{erg\,s^{-1}\,Hz^{-1}}} \right).
\end{multline}
Assuming an electron temperature of $T_e = 1 \times 10^4~\mathrm{K}$ and using the observed continuum luminosity, we derive 
$Q(\mathrm{H}^0) = 1.7 \times 10^{52}~\mathrm{s}^{-1}$.

Using the Starburst99 model \citep{Leitherer1999ApJS..123....3L} with a Kroupa initial mass function and assuming continuous star formation over 1~Myr, the observed ionizing photon rate can be translated into the SSC stellar mass, following the calibration presented by \citet{Leroy2018ApJ...869..126L}:
\begin{equation}
M_{\rm star} = \frac{Q_0}{10^{46.6}}M_\odot = 5.7 \times 10^5~M_\odot.
\end{equation}
This estimate is consistent with the stellar masses of the optically confirmed SSCs labeled in the light panel of Figure~\ref{fig:green}, supporting the interpretation that the ALMA 108~GHz continuum emission traces ionizing radiation from the same young stellar population identified in the HST-based catalog \citet{Whitmore2010}.

\subsection{Infrared view}

\subsubsection{Observational facts}

\begin{figure*}[ht!]
\centering
\includegraphics[scale=0.26]{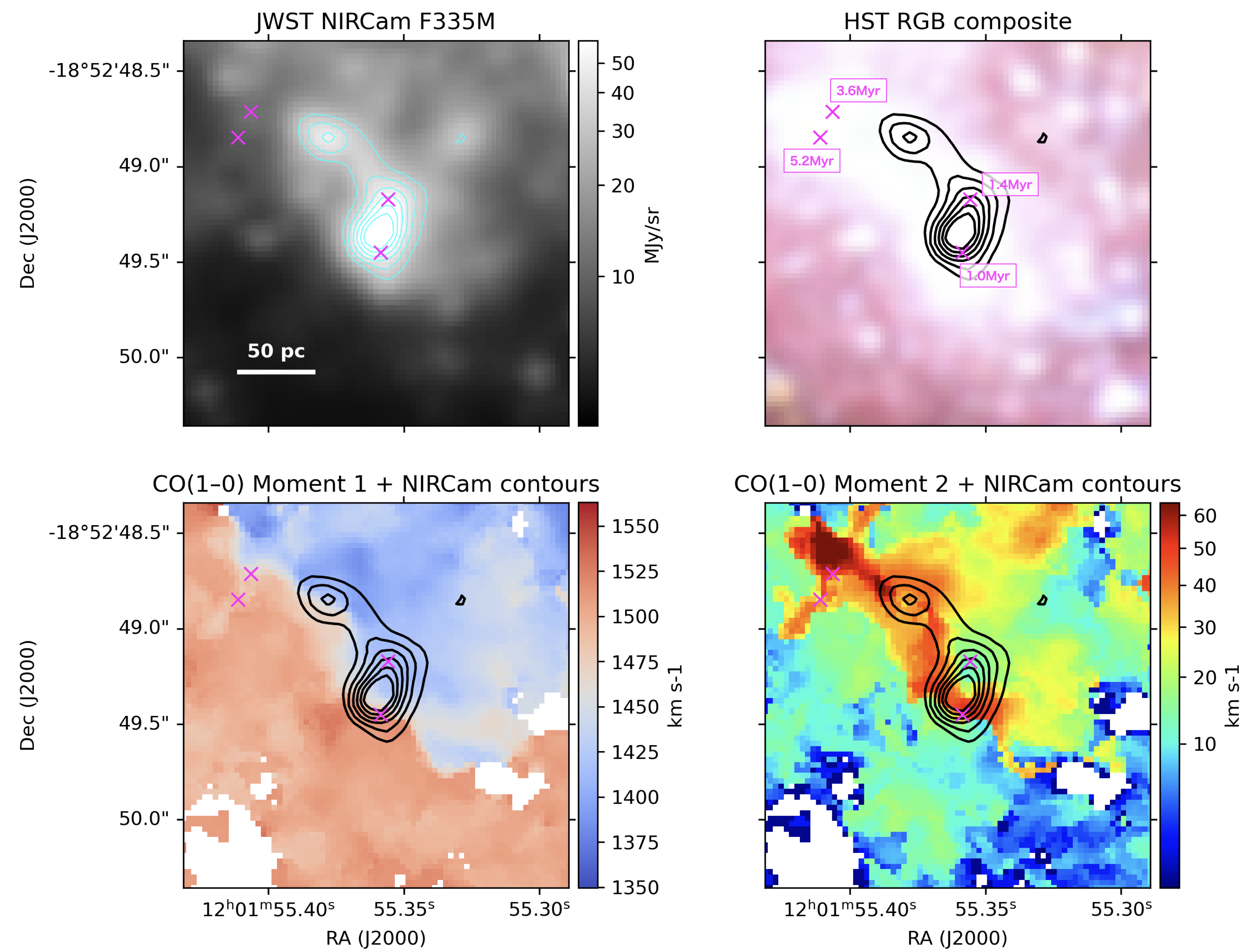} 
\caption{
{Multiwavelength comparison of the SGMC1–ALMA-3 region in the Antennae overlap.
Each panel covers a $2\arcsec\times2\arcsec$ field centered on the peak of the ALMA 108~GHz continuum emission.
Magenta crosses mark the positions of young massive star clusters identified in HST catalogs (Table~\ref{tab:SSCs}).
(Top left) JWST NIRCam F335M map shown in grayscale with cyan contours.
A 50~pc scale bar is shown for reference.
(Top right) HST optical RGB composite image (constructed from archival data) overlaid with the same NIRCam contours.
The HST and JWST images were manually aligned using stars, but a small systematic offset may still exist.
(Bottom left) ALMA CO~(1--0) moment 1 map with NIRCam contours, showing the kinematic distribution of the molecular gas relative to the NIRCam F335M emission.
(Bottom right) CO~(1--0) moment 2 (velocity-dispersion) map with NIRCam contours, emphasizing regions of enhanced velocity dispersion at the interface between gas components.}
\label{fig:comp}}
\end{figure*}

The HST cluster catalog \citep{Whitmore2010} identifies four star clusters within the field of interest: two very young clusters (IDs~19330 and 19275, with $\log_{10} \tau_{\rm cl} = 6.02$ and $6.16$) and two relatively older clusters (IDs~19459 and 19416, with $\log_{10} \tau_{\rm cl} = 6.56$ and $6.72$), where $\tau_{\rm cl}$ denotes the cluster age in years (see Table~\ref{tab:SSCs}).
As shown in Figure~\ref{fig:comp}, the JWST/NIRCam F335M emission, together with the MIRI F770M emission (Figure~\ref{fig:panel}), peaks at the locations of the younger clusters and coincides with the 108~GHz continuum peak, whereas the older clusters are modestly offset toward the west.

In addition, the F335M emission is distributed along the ridge of the redshifted CO filament.
Along this ridge, the CO line width is enhanced, and the F335M emission is preferentially located in these regions (Figure~\ref{fig:comp}).

\subsubsection{Physical interpretation}
Recent JWST studies have demonstrated that compact F335M sources can preferentially trace very young ($\lesssim3$~Myr) stellar clusters \citep{Rodriguez2023,Rodriguez2025}.
Our results do not contradict this trend, as the F335M emission in the present region is primarily associated with the younger clusters identified by HST.

The spatial distribution of the stellar clusters reveals a systematic offset, with older clusters located toward the northeastern side of the region and younger clusters concentrated near the 100~GHz continuum peak at the center.
Together with the alignment of the F335M emission along the redshifted molecular ridge and its association with regions of large CO velocity dispersion, these properties suggest that star formation has proceeded sequentially from the northeastern to the southwestern side along the ridge.

One plausible interpretation is that successive molecular cloud collisions along the ridge have contributed to the observed age distribution of stellar clusters and to the enhanced CO velocity dispersion.

\subsubsection{Limitations and outlook}

In addition to the discussion above, an important issue concerns the physical origin of the NIRCam F335M emission, specifically whether it predominantly traces the 3.3~$\mu$m PAH feature.
Addressing this issue in detail is beyond the scope of the present study.
Establishing the presence of a genuine 3.3~$\mu$m PAH excess requires a quantitative comparison between F335M and adjacent continuum-dominated filters, such as F300M and F360M, which would enable the separation of PAH emission from stellar continuum, particularly in compact clusters versus diffuse regions.

The present analysis therefore relies primarily on the spatial correspondence amoung F335M emission, gas kinematics, and the relative ages of stellar clusters inferred from existing HST-based catalogs.
A natural next step is to exploit the full capabilities of JWST by combining multi-band NIRCam flux ratios with physically motivated models of PAH excitation and destruction, together with the thermal and chemical evolution of shocked photodissociation interfaces \citep[e.g.,][]{Rigopoulou2021, Sidhu2022}.

In this context, it is plausible that PAH emission may be influenced not only by stellar feedback, which is commonly invoked to explain dust fragmentation \citep{Narayanan2023}, but also by shocks associated with molecular cloud collisions.
Disentangling these processes requires a joint treatment of PAH excitation, destruction, and grain fragmentation in dynamically evolving molecular environments.

By jointly analyzing PAH emission, gas dynamics, and JWST-based cluster properties, future studies will be able to place the local signatures of cloud--cloud collisions within a self-consistent evolutionary framework.
Extending such analyses to larger spatial scales will further link galaxy-scale gas flows, molecular cloud collisions, and PAH microphysics, offering a framework for interpreting collision-driven star formation.

\section{Conclusion}
\begin{figure*}[ht!]
\centering
\includegraphics[scale=0.24]{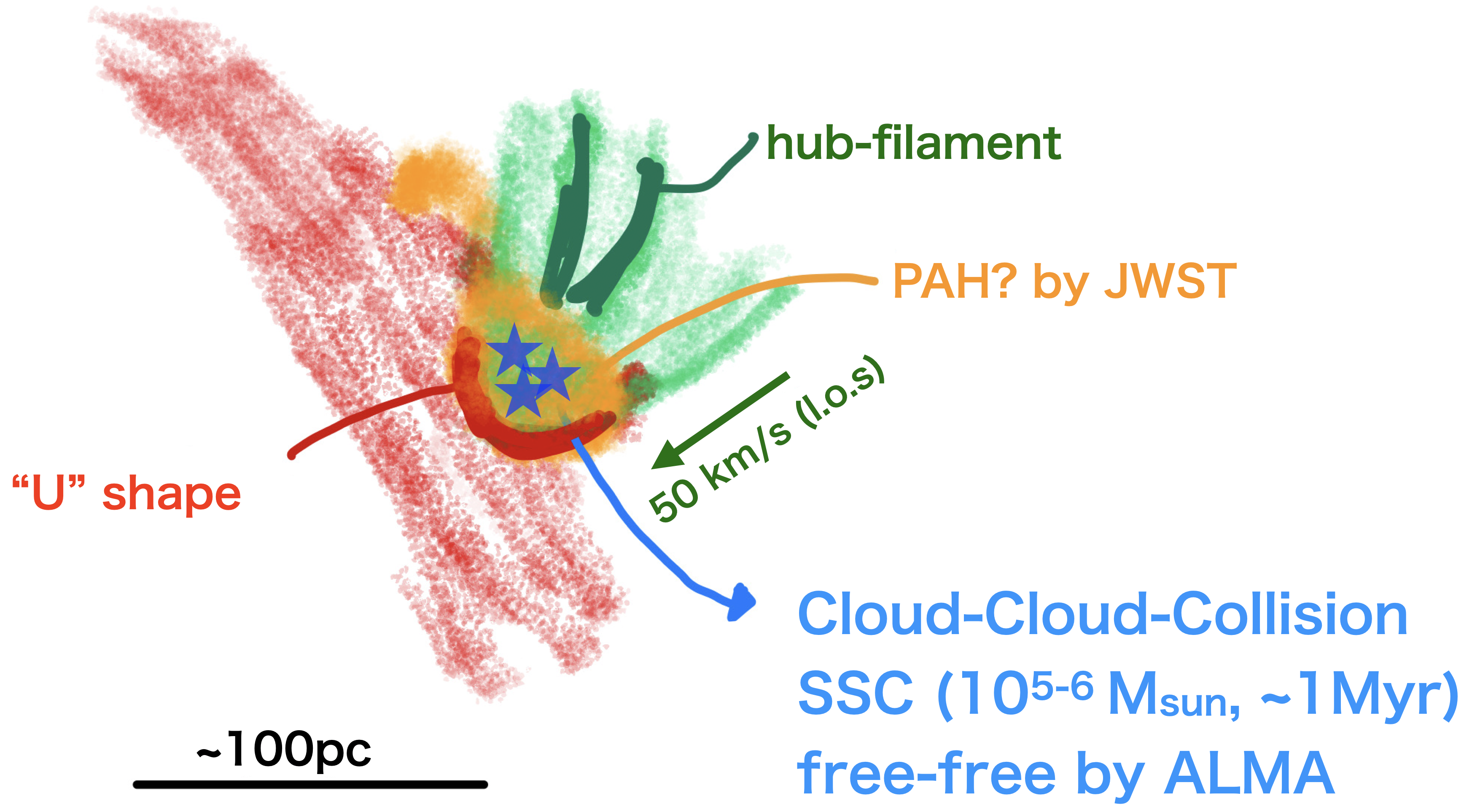} 
\caption{
Summary schematic of this study.
\label{fig:conclusion}}
\end{figure*}

Figure~\ref{fig:conclusion} schematically summarizes our findings.
Based on ALMA and JWST observations, we provide evidence that CCCs play a crucial role in triggering SSC formation within a super giant molecular cloud (SGMC1-ALMA-3) in the Antennae galaxies.
SGMC1-ALMA-3 is a particularly intriguing region, comparable to other notable SSC-forming sites such as the Firecracker \citep{Finn2019} and B1 \citep{Tsuge2021_B1}.
Key findings include:
\begin{itemize}
    \item High-resolution ALMA CO~($J=1$--$0$) observations of SGMC1-ALMA-3 reveal three key features that are naturally explained by a CCC model: a ``U"-shaped structure within a large filament, a hub–filament morphology, and a bridge-like feature in the PV diagram.
    \item  The ALMA 108~GHz continuum emission can be explained as free–free radiation from young massive stars, and it is spatially coincident with the JWST NIRCam F335M emission.
\end{itemize}

This study highlights the importance of CCCs in SSC formation and opens new avenues for future research. Follow-up investigations could include refining SSC properties using JWST data, developing more detailed theoretical models, and quantifying CCC activity across a broader area in the Antennae galaxies. \\

\begin{acknowledgments}
T.M. is supported by JSPS KAKENHI grant No. 25K17441.
This paper makes use of the following ALMA data: \texttt{ADS/JAO.ALMA\#2022.A.00032.S}.
ALMA is a partnership of ESO (representing its member states), NSF (USA) and NINS (Japan), together with NRC (Canada), NSTC and ASIAA (Taiwan), and KASI (Republic of Korea), in cooperation with the Republic of Chile. The Joint ALMA Observatory is operated by ESO, AUI/NRAO and NAOJ.
Some of the ALMA data were retrieved from the JVO portal (\url{http://jvo.nao.ac.jp/portal/}) operated by ADC/NAOJ. Data analysis was in part carried out on the common use data analysis computer system at the Astronomy Data Center, ADC, of the National Astronomical Observatory of Japan. 
{S.K. is supported by JSPS KAKENHI grant No. 25K07371.}
\end{acknowledgments}

\begin{contribution}
T.M. led the overall research, performed the analysis, and was responsible for writing the manuscript.  
T.S. and K.N. conducted the ALMA data reduction and contributed to the interpretation of the results.  
All authors provided scientific comments throughout the project and contributed to the revision of the manuscript.


\end{contribution}

%
\facilities{ALMA, JWST}

\software{
NumPy \citep{Harris2020Natur.585..357H},
SciPy \citep{2020SciPy-NMeth},
Matplotlib \citep{Hunter:2007},
Astropy \citep{2022ApJ...935..167A,2018AJ....156..123A,2013A&A...558A..33A},
ALMA Calibration Pipeline,
and CASA \citep{McMullin_2007ASPC..376..127M,CASA2022arXiv221002276T}.
}

\onecolumngrid

\twocolumngrid

\clearpage
\bibliography{sample7}{}

@ARTICLE{Sutter2024,
       author = {{Sutter}, Jessica and {Sandstrom}, Karin and {Chastenet}, J{\'e}r{\'e}my and {Leroy}, Adam K. and {Koch}, Eric W. and {Williams}, Thomas G. and {Chown}, Ryan and {Belfiore}, Francesco and {Bigiel}, Frank and {Boquien}, M{\'e}d{\'e}ric and {Cao}, Yixian and {Chevance}, M{\'e}lanie and {Dale}, Daniel A. and {Egorov}, Oleg V. and {Glover}, Simon C.~O. and {Groves}, Brent and {Klessen}, Ralf S. and {Kreckel}, Kathryn and {Larson}, Kirsten L. and {Oakes}, Elias K. and {Pathak}, Debosmita and {Ramambason}, Lise and {Rosolowsky}, Erik and {Watkins}, Elizabeth J.},
        title = "{The Fraction of Dust Mass in the Form of Polycyclic Aromatic Hydrocarbons on 10{\textendash}50 pc Scales in Nearby Galaxies}",
      journal = {\apj},
         year = 2024,
        month = aug,
       volume = {971},
       number = {2},
          eid = {178},
        pages = {178},
          doi = {10.3847/1538-4357/ad54bd},
archivePrefix = {arXiv},
       eprint = {2405.15102},
 primaryClass = {astro-ph.GA},
       adsurl = {https://ui.cadsabs.harvard.edu/abs/2024ApJ...971..178S}
}

@ARTICLE{Brunetti2024,
       author = {{Brunetti}, Nathan and {Wilson}, Christine D. and {He}, Hao and {Sun}, Jiayi and {Leroy}, Adam K. and {Rosolowsky}, Erik and {Bemis}, Ashley and {Bigiel}, Frank and {Groves}, Brent and {Saito}, Toshiki and {Schinnerer}, Eva},
        title = "{Cloud-scale molecular gas properties of the ANTENNAE merger: a comparative study with PHANGS-ALMA galaxies and NGC 3256}",
      journal = {\mnras},
         year = 2024,
        month = may,
       volume = {530},
       number = {1},
        pages = {597-612},
          doi = {10.1093/mnras/stae890},
archivePrefix = {arXiv},
       eprint = {2404.04555},
 primaryClass = {astro-ph.GA},
       adsurl = {https://ui.adsabs.harvard.edu/abs/2024MNRAS.530..597B}
}

@ARTICLE{Brunetti2022,
       author = {{Brunetti}, Nathan and {Wilson}, Christine D.},
        title = "{Extreme giant molecular clouds in the luminous infrared galaxy NGC 3256}",
      journal = {\mnras},
         year = 2022,
        month = sep,
       volume = {515},
       number = {2},
        pages = {2928-2950},
          doi = {10.1093/mnras/stac1975},
archivePrefix = {arXiv},
       eprint = {2207.05174},
 primaryClass = {astro-ph.GA},
       adsurl = {https://ui.adsabs.harvard.edu/abs/2022MNRAS.515.2928B}
}

@ARTICLE{Tokuda2022,
       author = {{Tokuda}, Kazuki and {Minami}, Taisei and {Fukui}, Yasuo and {Inoue}, Tsuyoshi and {Nishioka}, Takeru and {Tsuge}, Kisetsu and {Zahorecz}, Sarolta and {Sano}, Hidetoshi and {Konishi}, Ayu and {Rosie Chen}, C.-H. and {Sewi{\l}o}, Marta and {Madden}, Suzanne C. and {Nayak}, Omnarayani and {Saigo}, Kazuya and {Nishimura}, Atsushi and {Tanaka}, Kei E.~I. and {Sawada}, Tsuyoshi and {Indebetouw}, Remy and {Tachihara}, Kengo and {Kawamura}, Akiko and {Onishi}, Toshikazu},
        title = "{An ALMA Study of the Massive Molecular Clump N159W-North in the Large Magellanic Cloud: A Possible Gas Flow Penetrating One of the Most Massive Protocluster Systems in the Local Group}",
      journal = {\apj},
         year = 2022,
        month = jul,
       volume = {933},
       number = {1},
          eid = {20},
        pages = {20},
          doi = {10.3847/1538-4357/ac6b3c},
archivePrefix = {arXiv},
       eprint = {2205.00113},
 primaryClass = {astro-ph.GA},
       adsurl = {https://ui.adsabs.harvard.edu/abs/2022ApJ...933...20T}
}

@ARTICLE{Tsuge2024PASJ,
       author = {{Tsuge}, Kisetsu and {Sano}, Hidetoshi and {Tachihara}, Kengo and {Bekki}, Kenji and {Tokuda}, Kazuki and {Inoue}, Tsuyoshi and {Mizuno}, Norikazu and {Kawamura}, Akiko and {Onishi}, Toshikazu and {Fukui}, Yasuo},
        title = "{High-mass star formation in the Large Magellanic Cloud triggered by colliding H I flows}",
      journal = {\pasj},
         year = 2024,
        month = aug,
       volume = {76},
       number = {4},
        pages = {589-615},
          doi = {10.1093/pasj/psae035},
archivePrefix = {arXiv},
       eprint = {2405.05046},
 primaryClass = {astro-ph.GA},
       adsurl = {https://ui.adsabs.harvard.edu/abs/2024PASJ...76..589T}
}

@ARTICLE{Fukui2017PASJ,
       author = {{Fukui}, Yasuo and {Tsuge}, Kisetsu and {Sano}, Hidetoshi and {Bekki}, Kenji and {Yozin}, Cameron and {Tachihara}, Kengo and {Inoue}, Tsuyoshi},
        title = "{Formation of the young massive cluster R136 triggered by tidally-driven colliding H i flows}",
      journal = {\pasj},
         year = 2017,
        month = jun,
       volume = {69},
       number = {3},
          eid = {L5},
        pages = {L5},
          doi = {10.1093/pasj/psx032},
archivePrefix = {arXiv},
       eprint = {1703.01075},
 primaryClass = {astro-ph.GA},
       adsurl = {https://ui.adsabs.harvard.edu/abs/2017PASJ...69L...5F}
}

@ARTICLE{Sakamoto2014,
       author = {{Sakamoto}, Kazushi and {Aalto}, Susanne and {Combes}, Francoise and {Evans}, Aaron and {Peck}, Alison},
        title = "{An Infrared-luminous Merger with Two Bipolar Molecular Outflows: ALMA and SMA Observations of NGC 3256}",
      journal = {\apj},
         year = 2014,
        month = dec,
       volume = {797},
       number = {2},
          eid = {90},
        pages = {90},
          doi = {10.1088/0004-637X/797/2/90},
archivePrefix = {arXiv},
       eprint = {1403.7117},
 primaryClass = {astro-ph.GA},
       adsurl = {https://ui.adsabs.harvard.edu/abs/2014ApJ...797...90S}
}

@ARTICLE{Michiyama2018,
       author = {{Michiyama}, Tomonari and {Iono}, Daisuke and {Sliwa}, Kazimierz and {Bolatto}, Alberto and {Nakanishi}, Kouichiro and {Ueda}, Junko and {Saito}, Toshiki and {Ando}, Misaki and {Yamashita}, Takuji and {Yun}, Min},
        title = "{ALMA Observations of HCN and HCO$^{+}$ Outflows in the Merging Galaxy NGC 3256}",
      journal = {\apj},
         year = 2018,
        month = dec,
       volume = {868},
       number = {2},
          eid = {95},
        pages = {95},
          doi = {10.3847/1538-4357/aae82a},
archivePrefix = {arXiv},
       eprint = {1810.04821},
 primaryClass = {astro-ph.GA},
       adsurl = {https://ui.adsabs.harvard.edu/abs/2018ApJ...868...95M}
}

@ARTICLE{Brunetti2021,
       author = {{Brunetti}, Nathan and {Wilson}, Christine D. and {Sliwa}, Kazimierz and {Schinnerer}, Eva and {Aalto}, Susanne and {Peck}, Alison B.},
        title = "{Highly turbulent gas on GMC scales in NGC 3256, the nearest luminous infrared galaxy}",
      journal = {\mnras},
         year = 2021,
        month = jan,
       volume = {500},
       number = {4},
        pages = {4730-4748},
          doi = {10.1093/mnras/staa3425},
archivePrefix = {arXiv},
       eprint = {2011.01250},
 primaryClass = {astro-ph.GA},
       adsurl = {https://ui.adsabs.harvard.edu/abs/2021MNRAS.500.4730B}
}

@ARTICLE{Bournaud2015,
       author = {{Bournaud}, F. and {Daddi}, E. and {Wei{\ss}}, A. and {Renaud}, F. and {Mastropietro}, C. and {Teyssier}, R.},
        title = "{Modeling CO emission from hydrodynamic simulations of nearby spirals, starbursting mergers, and high-redshift galaxies}",
      journal = {\aap},
         year = 2015,
        month = mar,
       volume = {575},
          eid = {A56},
        pages = {A56},
          doi = {10.1051/0004-6361/201425078},
archivePrefix = {arXiv},
       eprint = {1409.8157},
 primaryClass = {astro-ph.GA},
       adsurl = {https://ui.adsabs.harvard.edu/abs/2015A&A...575A..56B}
}

@ARTICLE{Sidhu2022,
       author = {{Sidhu}, Ameek and {Tielens}, A.~G.~G.~M. and {Peeters}, Els and {Cami}, Jan},
        title = "{Polycyclic Aromatic Hydrocarbon emission model in photodissociation regions - I. Application to the 3.3, 6.2, and 11.2 {\ensuremath{\mu}}m bands}",
      journal = {\mnras},
         year = 2022,
        month = jul,
       volume = {514},
       number = {1},
        pages = {342-369},
          doi = {10.1093/mnras/stac1255},
archivePrefix = {arXiv},
       eprint = {2205.03304},
 primaryClass = {astro-ph.GA},
       adsurl = {https://ui.adsabs.harvard.edu/abs/2022MNRAS.514..342S}
}

@ARTICLE{Rigopoulou2021,
       author = {{Rigopoulou}, D. and {Barale}, M. and {Clary}, D.~C. and {Shan}, X. and {Alonso-Herrero}, A. and {Garc{\'\i}a-Bernete}, I. and {Hunt}, L. and {Kerkeni}, B. and {Pereira-Santaella}, M. and {Roche}, P.~F.},
        title = "{The properties of polycyclic aromatic hydrocarbons in galaxies: constraints on PAH sizes, charge and radiation fields}",
      journal = {\mnras},
         year = 2021,
        month = jul,
       volume = {504},
       number = {4},
        pages = {5287-5300},
          doi = {10.1093/mnras/stab959},
archivePrefix = {arXiv},
       eprint = {2011.10114},
 primaryClass = {astro-ph.GA},
       adsurl = {https://ui.adsabs.harvard.edu/abs/2021MNRAS.504.5287R}
}

@ARTICLE{Rodriguez2025,
       author = {{Rodr{\'\i}guez}, M. Jimena and {Lee}, Janice C. and {Indebetouw}, Remy and {Whitmore}, B.~C. and {Maschmann}, Daniel and {Williams}, Thomas G. and {Chandar}, Rupali and {Barnes}, A.~T. and {Gnedin}, Oleg Y. and {Sandstrom}, Karin M. and {Rosolowsky}, Erik and {Leroy}, Adam K. and {Thilker}, David A. and {Kim}, Hwihyun and {Sun}, Jiayi and {Klessen}, Ralf S. and {Groves}, Brent and {Wofford}, Aida and {Boquien}, M{\'e}d{\'e}ric and {Dale}, Daniel A. and {{\'U}beda}, Leonardo and {Larson}, Kirsten L. and {Grasha}, Kathryn and {Johnson}, Kelsey E. and {Levy}, Rebecca C. and {Bigiel}, Frank and {Hassani}, Hamid and {Sarbadhicary}, Sumit K.},
        title = "{Tracing the Earliest Stages of Star and Cluster Formation in 19 Nearby Galaxies with PHANGS-JWST and HST: Compact 3.3 {\ensuremath{\mu}}m Polycyclic Aromatic Hydrocarbon Emitters and Their Relation to the Optical Census of Star Clusters}",
      journal = {\apj},
         year = 2025,
        month = apr,
       volume = {983},
       number = {2},
          eid = {137},
        pages = {137},
          doi = {10.3847/1538-4357/adbb69},
archivePrefix = {arXiv},
       eprint = {2412.07862},
 primaryClass = {astro-ph.GA},
       adsurl = {https://ui.adsabs.harvard.edu/abs/2025ApJ...983..137R}
}

@ARTICLE{Narayanan2023,
       author = {{Narayanan}, Desika and {Smith}, J.-D.~T. and {Hensley}, Brandon S. and {Li}, Qi and {Hu}, Chia-Yu and {Sandstrom}, Karin and {Torrey}, Paul and {Vogelsberger}, Mark and {Marinacci}, Federico and {Sales}, Laura V.},
        title = "{A Framework for Modeling Polycyclic Aromatic Hydrocarbon Emission in Galaxy Evolution Simulations}",
      journal = {\apj},
         year = 2023,
        month = jul,
       volume = {951},
       number = {2},
          eid = {100},
        pages = {100},
          doi = {10.3847/1538-4357/accf8d},
archivePrefix = {arXiv},
       eprint = {2301.07136},
 primaryClass = {astro-ph.GA},
       adsurl = {https://ui.adsabs.harvard.edu/abs/2023ApJ...951..100N}
}

@ARTICLE{Whitmore1999,
       author = {{Whitmore}, Bradley C. and {Zhang}, Qing and {Leitherer}, Claus and {Fall}, S. Michael and {Schweizer}, Fran{\c{c}}ois and {Miller}, Bryan W.},
        title = "{The Luminosity Function of Young Star Clusters in ``the Antennae'' Galaxies (NGC 4038-4039)}",
      journal = {\aj},
         year = 1999,
        month = oct,
       volume = {118},
       number = {4},
        pages = {1551-1576},
          doi = {10.1086/301041},
archivePrefix = {arXiv},
       eprint = {astro-ph/9907430},
 primaryClass = {astro-ph},
       adsurl = {https://ui.adsabs.harvard.edu/abs/1999AJ....118.1551W}
}

@ARTICLE{Whitmore1995,
       author = {{Whitmore}, Bradley C. and {Schweizer}, Francois},
        title = "{Hubble Space Telescope Observations of Young Star Clusters in NGC 4038/4039, ``The Antennae'' Galaxies}",
      journal = {\aj},
         year = 1995,
        month = mar,
       volume = {109},
        pages = {960},
          doi = {10.1086/117334},
       adsurl = {https://ui.adsabs.harvard.edu/abs/1995AJ....109..960W}
}

@ARTICLE{Tasker2009,
       author = {{Tasker}, Elizabeth J. and {Tan}, Jonathan C.},
        title = "{Star Formation in Disk Galaxies. I. Formation and Evolution of Giant Molecular Clouds via Gravitational Instability and Cloud Collisions}",
      journal = {\apj},
         year = 2009,
        month = jul,
       volume = {700},
       number = {1},
        pages = {358-375},
          doi = {10.1088/0004-637X/700/1/358},
archivePrefix = {arXiv},
       eprint = {0811.0207},
 primaryClass = {astro-ph},
       adsurl = {https://ui.adsabs.harvard.edu/abs/2009ApJ...700..358T}
}

@ARTICLE{Tan2000,
       author = {{Tan}, Jonathan C.},
        title = "{Star Formation Rates in Disk Galaxies and Circumnuclear Starbursts from Cloud Collisions}",
      journal = {\apj},
         year = 2000,
        month = jun,
       volume = {536},
       number = {1},
        pages = {173-184},
          doi = {10.1086/308905},
archivePrefix = {arXiv},
       eprint = {astro-ph/9906355},
 primaryClass = {astro-ph},
       adsurl = {https://ui.adsabs.harvard.edu/abs/2000ApJ...536..173T}
}

@ARTICLE{Jog1992,
       author = {{Jog}, Chanda J. and {Das}, Mousumi},
        title = "{Starbursts Triggered by Central Overpressure in Interacting Galaxies}",
      journal = {\apj},
         year = 1992,
        month = dec,
       volume = {400},
        pages = {476},
          doi = {10.1086/172010},
       adsurl = {https://ui.adsabs.harvard.edu/abs/1992ApJ...400..476J}
}

@ARTICLE{Renaud2014,
       author = {{Renaud}, F. and {Bournaud}, F. and {Kraljic}, K. and {Duc}, P.-A.},
        title = "{Starbursts triggered by intergalactic tides andinterstellar compressive turbulence.}",
      journal = {\mnras},
         year = 2014,
        month = jul,
       volume = {442},
        pages = {L33-L37},
          doi = {10.1093/mnrasl/slu050},
archivePrefix = {arXiv},
       eprint = {1403.7316},
 primaryClass = {astro-ph.GA},
       adsurl = {https://ui.adsabs.harvard.edu/abs/2014MNRAS.442L..33R}
}

@ARTICLE{Renaud2019,
       author = {{Renaud}, F. and {Bournaud}, F. and {Agertz}, O. and {Kraljic}, K. and {Schinnerer}, E. and {Bolatto}, A. and {Daddi}, E. and {Hughes}, A.},
        title = "{A diversity of starburst-triggering mechanisms in interacting galaxies and their signatures in CO emission}",
      journal = {\aap},
         year = 2019,
        month = may,
       volume = {625},
          eid = {A65},
        pages = {A65},
          doi = {10.1051/0004-6361/201935222},
archivePrefix = {arXiv},
       eprint = {1902.02353},
 primaryClass = {astro-ph.GA},
       adsurl = {https://ui.adsabs.harvard.edu/abs/2019A&A...625A..65R}
}

@ARTICLE{Rodriguez2023,
       author = {{Rodr{\'\i}guez}, M. Jimena and {Lee}, Janice C. and {Whitmore}, B.~C. and {Thilker}, David A. and {Maschmann}, Daniel and {Chandar}, Rupali and {Deger}, Sinan and {Boquien}, M{\'e}d{\'e}ric and {Dale}, Daniel A. and {Larson}, Kirsten L. and {Williams}, Thomas G. and {Kim}, Hwihyun and {Schinnerer}, Eva and {Rosolowsky}, Erik and {Leroy}, Adam K. and {Emsellem}, Eric and {Sandstrom}, Karin M. and {Kruijssen}, J.~M. Diederik and {Grasha}, Kathryn and {Watkins}, Elizabeth J. and {Barnes}, Ashley. T. and {Sormani}, Mattia C. and {Kim}, Jaeyeon and {Anand}, Gagandeep S. and {Chevance}, M{\'e}lanie and {Bigiel}, F. and {Klessen}, Ralf S. and {Hassani}, Hamid and {Liu}, Daizhong and {Faesi}, Christopher M. and {Cao}, Yixian and {Belfiore}, Francesco and {Pessa}, Ismael and {Kreckel}, Kathryn and {Groves}, Brent and {Pety}, J{\'e}r{\^o}me and {Indebetouw}, R{\'e}my and {Egorov}, Oleg V. and {Blanc}, Guillermo A. and {Saito}, Toshiki and {Hughes}, Annie},
        title = "{PHANGS-JWST First Results: Dust-embedded Star Clusters in NGC 7496 Selected via 3.3 {\ensuremath{\mu}}m PAH Emission}",
      journal = {\apjl},
         year = 2023,
        month = feb,
       volume = {944},
       number = {2},
          eid = {L26},
        pages = {L26},
          doi = {10.3847/2041-8213/aca653},
archivePrefix = {arXiv},
       eprint = {2211.13426},
 primaryClass = {astro-ph.GA},
       adsurl = {https://ui.adsabs.harvard.edu/abs/2023ApJ...944L..26R}
}

@ARTICLE{Bolatto2013,
       author = {{Bolatto}, Alberto D. and {Wolfire}, Mark and {Leroy}, Adam K.},
        title = "{The CO-to-H$_{2}$ Conversion Factor}",
      journal = {\araa},
         year = 2013,
        month = aug,
       volume = {51},
       number = {1},
        pages = {207-268},
          doi = {10.1146/annurev-astro-082812-140944},
archivePrefix = {arXiv},
       eprint = {1301.3498},
 primaryClass = {astro-ph.GA},
       adsurl = {https://ui.adsabs.harvard.edu/abs/2013ARA&A..51..207B}
}

@article{He22,
doi = {10.3847/1538-4357/ac5628},
url = {https://dx.doi.org/10.3847/1538-4357/ac5628},
year = {2022},
month = {mar},
publisher = {The American Astronomical Society},
volume = {928},
number = {1},
pages = {57},
author = {He, Hao and Wilson, Christine and Brunetti, Nathan and Finn, Molly and Bemis, Ashley and Johnson, Kelsey},
title = {Embedded Young Massive Star Clusters in the Antennae Merger},
journal = {The Astrophysical Journal}
}

@ARTICLE{Rubin1968,
       author = {{Rubin}, Robert H.},
        title = "{A Discussion of the Sizes and Excitation of H II Regions}",
      journal = {\apj},
         year = 1968,
        month = oct,
       volume = {154},
        pages = {391},
          doi = {10.1086/149766},
       adsurl = {https://ui.adsabs.harvard.edu/abs/1968ApJ...154..391R}
}

@ARTICLE{Harris2020Natur.585..357H,
       author = {{Harris}, Charles R. and {Millman}, K. Jarrod and {van der Walt}, St{\'e}fan J. and {Gommers}, Ralf and {Virtanen}, Pauli and {Cournapeau}, David and {Wieser}, Eric and {Taylor}, Julian and {Berg}, Sebastian and {Smith}, Nathaniel J. and {Kern}, Robert and {Picus}, Matti and {Hoyer}, Stephan and {van Kerkwijk}, Marten H. and {Brett}, Matthew and {Haldane}, Allan and {del R{\'\i}o}, Jaime Fern{\'a}ndez and {Wiebe}, Mark and {Peterson}, Pearu and {G{\'e}rard-Marchant}, Pierre and {Sheppard}, Kevin and {Reddy}, Tyler and {Weckesser}, Warren and {Abbasi}, Hameer and {Gohlke}, Christoph and {Oliphant}, Travis E.},
        title = "{Array programming with NumPy}",
      journal = {\nat},
         year = 2020,
        month = sep,
       volume = {585},
       number = {7825},
        pages = {357-362},
          doi = {10.1038/s41586-020-2649-2},
archivePrefix = {arXiv},
       eprint = {2006.10256},
 primaryClass = {cs.MS},
       adsurl = {https://ui.adsabs.harvard.edu/abs/2020Natur.585..357H}
}

@Article{Hunter:2007,
  Author    = {Hunter, J. D.},
  Title     = {Matplotlib: A 2D graphics environment},
  Journal   = {Computing in Science \& Engineering},
  Volume    = {9},
  Number    = {3},
  Pages     = {90--95},
  publisher = {IEEE COMPUTER SOC},
  doi       = {10.1109/MCSE.2007.55},
  year      = 2007
}

@INPROCEEDINGS{McMullin_2007ASPC..376..127M,
       author = {{McMullin}, J.~P. and {Waters}, B. and {Schiebel}, D. and {Young}, W. and {Golap}, K.},
        title = "{CASA Architecture and Applications}",
    booktitle = {Astronomical Data Analysis Software and Systems XVI},
         year = 2007,
       editor = {{Shaw}, R.~A. and {Hill}, F. and {Bell}, D.~J.},
       series = {Astronomical Society of the Pacific Conference Series},
       volume = {376},
        month = oct,
        pages = {127},
       adsurl = {https://ui.adsabs.harvard.edu/abs/2007ASPC..376..127M}
}

@ARTICLE{CASA2022arXiv221002276T,
       author = {{THE CASA TEAM} and {Bean}, Ben and {Bhatnagar}, Sanjay and {Castro}, Sandra and {Donovan Meyer}, Jennifer and {Emonts}, Bjorn and {Garcia}, Enrique and {Garwood}, Robert and {Golap}, Kumar and {Gonzalez Villalba}, Justo and {Harris}, Pamela and {Hayashi}, Yohei and {Hoskins}, Josh and {Hsieh}, Mingyu and {Jagannathan}, Preshanth and {Kawasaki}, Wataru and {Keimpema}, Aard and {Kettenis}, Mark and {Lopez}, Jorge and {Marvil}, Joshua and {Masters}, Joseph and {McNichols}, Andrew and {Mehringer}, David and {Miel}, Renaud and {Moellenbrock}, George and {Montesino}, Federico and {Nakazato}, Takeshi and {Ott}, Juergen and {Petry}, Dirk and {Pokorny}, Martin and {Raba}, Ryan and {Rau}, Urvashi and {Schiebel}, Darrell and {Schweighart}, Neal and {Sekhar}, Srikrishna and {Shimada}, Kazuhiko and {Small}, Des and {Steeb}, Jan-Willem and {Sugimoto}, Kanako and {Suoranta}, Ville and {Tsutsumi}, Takahiro and {van Bemmel}, Ilse M. and {Verkouter}, Marjolein and {Wells}, Akeem and {Xiong}, Wei and {Szomoru}, Arpad and {Griffith}, Morgan and {Glendenning}, Brian and {Kern}, Jeff},
        title = "{CASA, the Common Astronomy Software Applications for Radio Astronomy}",
      journal = {arXiv e-prints},
         year = 2022,
        month = oct,
          eid = {arXiv:2210.02276},
        pages = {arXiv:2210.02276},
archivePrefix = {arXiv},
       eprint = {2210.02276},
 primaryClass = {astro-ph.IM},
       adsurl = {https://ui.adsabs.harvard.edu/abs/2022arXiv221002276T}
}

@ARTICLE{2020SciPy-NMeth,
  author  = {Virtanen, Pauli and Gommers, Ralf and Oliphant, Travis E. and Haberland, Matt and Reddy, Tyler and Cournapeau, David and Burovski, Evgeni and Peterson, Pearu and Weckesser, Warren and Bright, Jonathan and {van der Walt}, St{\'e}fan J. and Brett, Matthew and Wilson, Joshua and Millman, K. Jarrod and Mayorov, Nikolay and Nelson, Andrew R. J. and Jones, Eric and Kern, Robert and Larson, Eric and Carey, C J and Polat, {\.I}lhan and Feng, Yu and Moore, Eric W. and {VanderPlas}, Jake and Laxalde, Denis and Perktold, Josef and Cimrman, Robert and Henriksen, Ian and Quintero, E. A. and Harris, Charles R. and Archibald, Anne M. and Ribeiro, Ant{\^o}nio H. and Pedregosa, Fabian and {van Mulbregt}, Paul and {SciPy 1.0 Contributors}},
  title   = {{{SciPy} 1.0: Fundamental Algorithms for Scientific
            Computing in Python}},
  journal = {Nature Methods},
  year    = {2020},
  volume  = {17},
  pages   = {261--272},
  adsurl  = {https://rdcu.be/b08Wh},
  doi     = {10.1038/s41592-019-0686-2},
}

@ARTICLE{Solomon1987,
       author = {{Solomon}, P.~M. and {Rivolo}, A.~R. and {Barrett}, J. and {Yahil}, A.},
        title = "{Mass, Luminosity, and Line Width Relations of Galactic Molecular Clouds}",
      journal = {\apj},
         year = 1987,
        month = aug,
       volume = {319},
        pages = {730},
          doi = {10.1086/165493},
       adsurl = {https://ui.adsabs.harvard.edu/abs/1987ApJ...319..730S}
}

@ARTICLE{Torii2017,
       author = {{Torii}, K. and {Hattori}, Y. and {Hasegawa}, K. and {Ohama}, A. and {Haworth}, T.~J. and {Shima}, K. and {Habe}, A. and {Tachihara}, K. and {Mizuno}, N. and {Onishi}, T. and {Mizuno}, A. and {Fukui}, Y.},
        title = "{Triggered O Star Formation in M20 via Cloud-Cloud Collision: Comparisons between High-resolution CO Observations and Simulations}",
      journal = {\apj},
         year = 2017,
        month = feb,
       volume = {835},
       number = {2},
          eid = {142},
        pages = {142},
          doi = {10.3847/1538-4357/835/2/142},
archivePrefix = {arXiv},
       eprint = {1612.09458},
 primaryClass = {astro-ph.GA},
       adsurl = {https://ui.adsabs.harvard.edu/abs/2017ApJ...835..142T}
}

@ARTICLE{Torii2015,
       author = {{Torii}, K. and {Hasegawa}, K. and {Hattori}, Y. and {Sano}, H. and {Ohama}, A. and {Yamamoto}, H. and {Tachihara}, K. and {Soga}, S. and {Shimizu}, S. and {Okuda}, T. and {Mizuno}, N. and {Onishi}, T. and {Mizuno}, A. and {Fukui}, Y.},
        title = "{Cloud-Cloud Collision as a Trigger of the High-mass Star Formation: a Molecular Line Study in RCW120}",
      journal = {\apj},
         year = 2015,
        month = jun,
       volume = {806},
       number = {1},
          eid = {7},
        pages = {7},
          doi = {10.1088/0004-637X/806/1/7},
archivePrefix = {arXiv},
       eprint = {1503.00070},
 primaryClass = {astro-ph.GA},
       adsurl = {https://ui.adsabs.harvard.edu/abs/2015ApJ...806....7T}
}

@ARTICLE{Haworth2015MNRAS.454.1634H,
       author = {{Haworth}, T.~J. and {Shima}, K. and {Tasker}, E.~J. and {Fukui}, Y. and {Torii}, K. and {Dale}, J.~E. and {Takahira}, K. and {Habe}, A.},
        title = "{Isolating signatures of major cloud-cloud collisions - II. The lifetimes of broad bridge features}",
      journal = {\mnras},
         year = 2015,
        month = dec,
       volume = {454},
       number = {2},
        pages = {1634-1643},
          doi = {10.1093/mnras/stv2068},
archivePrefix = {arXiv},
       eprint = {1509.00859},
 primaryClass = {astro-ph.GA},
       adsurl = {https://ui.adsabs.harvard.edu/abs/2015MNRAS.454.1634H}
}

@ARTICLE{Haworth2015MNRAS.450...10H,
       author = {{Haworth}, T.~J. and {Tasker}, E.~J. and {Fukui}, Y. and {Torii}, K. and {Dale}, J.~E. and {Shima}, K. and {Takahira}, K. and {Habe}, A. and {Hasegawa}, K.},
        title = "{Isolating signatures of major cloud-cloud collisions using position-velocity diagrams}",
      journal = {\mnras},
         year = 2015,
        month = jun,
       volume = {450},
       number = {1},
        pages = {10-20},
          doi = {10.1093/mnras/stv639},
archivePrefix = {arXiv},
       eprint = {1503.06795},
 primaryClass = {astro-ph.GA},
       adsurl = {https://ui.adsabs.harvard.edu/abs/2015MNRAS.450...10H}
}

@ARTICLE{Takahira2014ApJ...792...63T,
       author = {{Takahira}, Ken and {Tasker}, Elizabeth J. and {Habe}, Asao},
        title = "{Do Cloud-Cloud Collisions Trigger High-mass Star Formation? I. Small Cloud Collisions}",
      journal = {\apj},
         year = 2014,
        month = sep,
       volume = {792},
       number = {1},
          eid = {63},
        pages = {63},
          doi = {10.1088/0004-637X/792/1/63},
archivePrefix = {arXiv},
       eprint = {1407.4544},
 primaryClass = {astro-ph.GA},
       adsurl = {https://ui.adsabs.harvard.edu/abs/2014ApJ...792...63T}
}

@ARTICLE{Leitherer1999ApJS..123....3L,
       author = {{Leitherer}, Claus and {Schaerer}, Daniel and {Goldader}, Jeffrey D. and {Delgado}, Rosa M. Gonz{\'a}lez and {Robert}, Carmelle and {Kune}, Denis Foo and {de Mello}, Du{\'\i}lia F. and {Devost}, Daniel and {Heckman}, Timothy M.},
        title = "{Starburst99: Synthesis Models for Galaxies with Active Star Formation}",
      journal = {\apjs},
         year = 1999,
        month = jul,
       volume = {123},
       number = {1},
        pages = {3-40},
          doi = {10.1086/313233},
archivePrefix = {arXiv},
       eprint = {astro-ph/9902334},
 primaryClass = {astro-ph},
       adsurl = {https://ui.adsabs.harvard.edu/abs/1999ApJS..123....3L}
}

@ARTICLE{Leroy2018ApJ...869..126L,
       author = {{Leroy}, Adam K. and {Bolatto}, Alberto D. and {Ostriker}, Eve C. and {Walter}, Fabian and {Gorski}, Mark and {Ginsburg}, Adam and {Krieger}, Nico and {Levy}, Rebecca C. and {Meier}, David S. and {Mills}, Elisabeth and {Ott}, J{\"u}rgen and {Rosolowsky}, Erik and {Thompson}, Todd A. and {Veilleux}, Sylvain and {Zschaechner}, Laura K.},
        title = "{Forming Super Star Clusters in the Central Starburst of NGC 253}",
      journal = {\apj},
         year = 2018,
        month = dec,
       volume = {869},
       number = {2},
          eid = {126},
        pages = {126},
          doi = {10.3847/1538-4357/aaecd1},
archivePrefix = {arXiv},
       eprint = {1804.02083},
 primaryClass = {astro-ph.GA},
       adsurl = {https://ui.adsabs.harvard.edu/abs/2018ApJ...869..126L}
}

@ARTICLE{Sewi_2023ApJ...959...22S,
       author = {{Sewi{\l}o}, Marta and {Tokuda}, Kazuki and {Kurtz}, Stan E. and {Charnley}, Steven B. and {M{\"o}ller}, Thomas and {Wiseman}, Jennifer and {Chen}, C. -H. Rosie and {Indebetouw}, Remy and {S{\'a}nchez-Monge}, {\'A}lvaro and {Tanaka}, Kei E.~I. and {Schilke}, Peter and {Onishi}, Toshikazu and {Harada}, Naoto},
        title = "{The Detection of Higher-order Millimeter Hydrogen Recombination Lines in the Large Magellanic Cloud}",
      journal = {\apj},
         year = 2023,
        month = dec,
       volume = {959},
       number = {1},
          eid = {22},
        pages = {22},
          doi = {10.3847/1538-4357/acf5ed},
archivePrefix = {arXiv},
       eprint = {2309.02586},
 primaryClass = {astro-ph.GA},
       adsurl = {https://ui.adsabs.harvard.edu/abs/2023ApJ...959...22S}
}

@ARTICLE{Tokuda2019ApJ...886...15T,
       author = {{Tokuda}, Kazuki and {Fukui}, Yasuo and {Harada}, Ryohei and {Saigo}, Kazuya and {Tachihara}, Kengo and {Tsuge}, Kisetsu and {Inoue}, Tsuyoshi and {Torii}, Kazufumi and {Nishimura}, Atsushi and {Zahorecz}, Sarolta and {Nayak}, Omnarayani and {Meixner}, Margaret and {Minamidani}, Tetsuhiro and {Kawamura}, Akiko and {Mizuno}, Norikazu and {Indebetouw}, Remy and {Sewi{\l}o}, Marta and {Madden}, Suzanne and {Galametz}, Maud and {Lebouteiller}, Vianney and {Chen}, C. -H. Rosie and {Onishi}, Toshikazu},
        title = "{An ALMA View of Molecular Filaments in the Large Magellanic Cloud. II. An Early Stage of High-mass Star Formation Embedded at Colliding Clouds in N159W-South}",
      journal = {\apj},
         year = 2019,
        month = nov,
       volume = {886},
       number = {1},
          eid = {15},
        pages = {15},
          doi = {10.3847/1538-4357/ab48ff},
archivePrefix = {arXiv},
       eprint = {1811.04400},
 primaryClass = {astro-ph.GA},
       adsurl = {https://ui.adsabs.harvard.edu/abs/2019ApJ...886...15T}
}

@ARTICLE{Fukui2019ApJ...886...14F,
       author = {{Fukui}, Yasuo and {Tokuda}, Kazuki and {Saigo}, Kazuya and {Harada}, Ryohei and {Tachihara}, Kengo and {Tsuge}, Kisetsu and {Inoue}, Tsuyoshi and {Torii}, Kazufumi and {Nishimura}, Atsushi and {Zahorecz}, Sarolta and {Nayak}, Omnarayani and {Meixner}, Margaret and {Minamidani}, Tetsuhiro and {Kawamura}, Akiko and {Mizuno}, Norikazu and {Indebetouw}, Remy and {Sewi{\l}o}, Marta and {Madden}, Suzanne and {Galametz}, Maud and {Lebouteiller}, Vianney and {Chen}, C. -H. Rosie and {Onishi}, Toshikazu},
        title = "{An ALMA View of Molecular Filaments in the Large Magellanic Cloud. I. The Formation of High-mass Stars and Pillars in the N159E-Papillon Nebula Triggered by a Cloud-Cloud Collision}",
      journal = {\apj},
         year = 2019,
        month = nov,
       volume = {886},
       number = {1},
          eid = {14},
        pages = {14},
          doi = {10.3847/1538-4357/ab4900},
archivePrefix = {arXiv},
       eprint = {1811.00812},
 primaryClass = {astro-ph.GA},
       adsurl = {https://ui.adsabs.harvard.edu/abs/2019ApJ...886...14F}
}

@ARTICLE{Maity2024ApJ...974..229M,
       author = {{Maity}, A.~K. and {Inoue}, T. and {Fukui}, Y. and {Dewangan}, L.~K. and {Sano}, H. and {Yamada}, R.~I. and {Tachihara}, K. and {Bhadari}, N.~K. and {Jadhav}, O.~R.},
        title = "{Cloud{\textendash}Cloud Collision: Formation of Hub-filament Systems and Associated Gas Kinematics. Mass-collecting Cone{\textemdash}A New Signature of Cloud{\textendash}Cloud Collision}",
      journal = {\apj},
         year = 2024,
        month = oct,
       volume = {974},
       number = {2},
          eid = {229},
        pages = {229},
          doi = {10.3847/1538-4357/ad7098},
archivePrefix = {arXiv},
       eprint = {2408.06826},
 primaryClass = {astro-ph.GA},
       adsurl = {https://ui.adsabs.harvard.edu/abs/2024ApJ...974..229M}
}

@ARTICLE{Inoue2018PASJ...70S..53I,
       author = {{Inoue}, Tsuyoshi and {Hennebelle}, Patrick and {Fukui}, Yasuo and {Matsumoto}, Tomoaki and {Iwasaki}, Kazunari and {Inutsuka}, Shu-ichiro},
        title = "{The formation of massive molecular filaments and massive stars triggered by a magnetohydrodynamic shock wave}",
      journal = {\pasj},
         year = 2018,
        month = may,
       volume = {70},
          eid = {S53},
        pages = {S53},
          doi = {10.1093/pasj/psx089},
archivePrefix = {arXiv},
       eprint = {1707.02035},
 primaryClass = {astro-ph.GA},
       adsurl = {https://ui.adsabs.harvard.edu/abs/2018PASJ...70S..53I}
}

@ARTICLE{Schinnerer2024ARA&A..62..369S,
       author = {{Schinnerer}, E. and {Leroy}, A.~K.},
        title = "{Molecular Gas and the Star-Formation Process on Cloud Scales in Nearby Galaxies}",
      journal = {\araa},
         year = 2024,
        month = sep,
       volume = {62},
       number = {1},
        pages = {369-436},
          doi = {10.1146/annurev-astro-071221-052651},
archivePrefix = {arXiv},
       eprint = {2403.19843},
 primaryClass = {astro-ph.GA},
       adsurl = {https://ui.adsabs.harvard.edu/abs/2024ARA&A..62..369S}
}

@ARTICLE{Murphy2011,
       author = {{Murphy}, E.~J. and {Condon}, J.~J. and {Schinnerer}, E. and {Kennicutt}, R.~C. and {Calzetti}, D. and {Armus}, L. and {Helou}, G. and {Turner}, J.~L. and {Aniano}, G. and {Beir{\~a}o}, P. and {Bolatto}, A.~D. and {Brandl}, B.~R. and {Croxall}, K.~V. and {Dale}, D.~A. and {Donovan Meyer}, J.~L. and {Draine}, B.~T. and {Engelbracht}, C. and {Hunt}, L.~K. and {Hao}, C. -N. and {Koda}, J. and {Roussel}, H. and {Skibba}, R. and {Smith}, J. -D.~T.},
        title = "{Calibrating Extinction-free Star Formation Rate Diagnostics with 33 GHz Free-free Emission in NGC 6946}",
      journal = {\apj},
         year = 2011,
        month = aug,
       volume = {737},
       number = {2},
          eid = {67},
        pages = {67},
          doi = {10.1088/0004-637X/737/2/67},
archivePrefix = {arXiv},
       eprint = {1105.4877},
 primaryClass = {astro-ph.CO},
       adsurl = {https://ui.adsabs.harvard.edu/abs/2011ApJ...737...67M}
}

@ARTICLE{Whitmore2014,
       author = {{Whitmore}, Bradley C. and {Brogan}, Crystal and {Chandar}, Rupali and {Evans}, Aaron and {Hibbard}, John and {Johnson}, Kelsey and {Leroy}, Adam and {Privon}, George and {Remijan}, Anthony and {Sheth}, Kartik},
        title = "{ALMA Observations of the Antennae Galaxies. I. A New Window on a Prototypical Merger}",
      journal = {\apj},
         year = 2014,
        month = nov,
       volume = {795},
       number = {2},
          eid = {156},
        pages = {156},
          doi = {10.1088/0004-637X/795/2/156},
archivePrefix = {arXiv},
       eprint = {1410.4473},
 primaryClass = {astro-ph.GA},
       adsurl = {https://ui.adsabs.harvard.edu/abs/2014ApJ...795..156W}
}

@ARTICLE{Fukui2014,
       author = {{Fukui}, Y. and {Ohama}, A. and {Hanaoka}, N. and {Furukawa}, N. and {Torii}, K. and {Dawson}, J.~R. and {Mizuno}, N. and {Hasegawa}, K. and {Fukuda}, T. and {Soga}, S. and {Moribe}, N. and {Kuroda}, Y. and {Hayakawa}, T. and {Kawamura}, A. and {Kuwahara}, T. and {Yamamoto}, H. and {Okuda}, T. and {Onishi}, T. and {Maezawa}, H. and {Mizuno}, A.},
        title = "{Molecular Clouds toward the Super Star Cluster NGC 3603 Possible Evidence for a Cloud-Cloud Collision in Triggering the Cluster Formation}",
      journal = {\apj},
         year = 2014,
        month = jan,
       volume = {780},
       number = {1},
          eid = {36},
        pages = {36},
          doi = {10.1088/0004-637X/780/1/36},
archivePrefix = {arXiv},
       eprint = {1306.2090},
 primaryClass = {astro-ph.GA},
       adsurl = {https://ui.adsabs.harvard.edu/abs/2014ApJ...780...36F}
}

@ARTICLE{Elmegreen1997,
       author = {{Elmegreen}, Bruce G. and {Efremov}, Yuri N.},
        title = "{A Universal Formation Mechanism for Open and Globular Clusters in Turbulent Gas}",
      journal = {\apj},
         year = 1997,
        month = may,
       volume = {480},
       number = {1},
        pages = {235-245},
          doi = {10.1086/303966},
       adsurl = {https://ui.adsabs.harvard.edu/abs/1997ApJ...480..235E}
}

@ARTICLE{Inoue2013,
       author = {{Inoue}, Tsuyoshi and {Fukui}, Yasuo},
        title = "{Formation of Massive Molecular Cloud Cores by Cloud-Cloud Collision}",
      journal = {\apjl},
         year = 2013,
        month = sep,
       volume = {774},
       number = {2},
          eid = {L31},
        pages = {L31},
          doi = {10.1088/2041-8205/774/2/L31},
archivePrefix = {arXiv},
       eprint = {1305.4655},
 primaryClass = {astro-ph.GA},
       adsurl = {https://ui.adsabs.harvard.edu/abs/2013ApJ...774L..31I}
}

@ARTICLE{Tsuge2021,
       author = {{Tsuge}, Kisetsu and {Fukui}, Yasuo and {Tachihara}, Kengo and {Sano}, Hidetoshi and {Tokuda}, Kazuki and {Ueda}, Junko and {Iono}, Daisuke and {Finn}, Molly K.},
        title = "{The formation of young massive clusters triggered by cloud-cloud collisions in the Antennae galaxies NGC 4038/NGC 4039}",
      journal = {\pasj},
         year = 2021,
        month = jan,
       volume = {73},
        pages = {S35-S61},
          doi = {10.1093/pasj/psaa033},
archivePrefix = {arXiv},
       eprint = {1909.05240},
 primaryClass = {astro-ph.GA},
       adsurl = {https://ui.adsabs.harvard.edu/abs/2021PASJ...73S..35T}
}

@ARTICLE{Fukui2021,
       author = {{Fukui}, Yasuo and {Habe}, Asao and {Inoue}, Tsuyoshi and {Enokiya}, Rei and {Tachihara}, Kengo},
        title = "{Cloud-cloud collisions and triggered star formation}",
      journal = {\pasj},
         year = 2021,
        month = jan,
       volume = {73},
        pages = {S1-S34},
          doi = {10.1093/pasj/psaa103},
archivePrefix = {arXiv},
       eprint = {2009.05077},
 primaryClass = {astro-ph.GA},
       adsurl = {https://ui.adsabs.harvard.edu/abs/2021PASJ...73S...1F}
}

@ARTICLE{Johnson2015,
       author = {{Johnson}, K.~E. and {Leroy}, A.~K. and {Indebetouw}, R. and {Brogan}, C.~L. and {Whitmore}, B.~C. and {Hibbard}, J. and {Sheth}, K. and {Evans}, A.~S.},
        title = "{The Physical Conditions in a Pre-super Star Cluster Molecular Cloud in the Antennae Galaxies}",
      journal = {\apj},
         year = 2015,
        month = jun,
       volume = {806},
       number = {1},
          eid = {35},
        pages = {35},
          doi = {10.1088/0004-637X/806/1/35},
archivePrefix = {arXiv},
       eprint = {1503.06477},
 primaryClass = {astro-ph.GA},
       adsurl = {https://ui.adsabs.harvard.edu/abs/2015ApJ...806...35J}
}

@ARTICLE{Tsuge2021_B1,
       author = {{Tsuge}, Kisetsu and {Tachihara}, Kengo and {Fukui}, Yasuo and {Sano}, Hidetoshi and {Tokuda}, Kazuki and {Ueda}, Junko and {Iono}, Daisuke},
        title = "{The formation of the young massive cluster B1 in the Antennae Galaxies (NGC 4038/NGC 4039) triggered by cloud-cloud collision}",
      journal = {\pasj},
         year = 2021,
        month = apr,
       volume = {73},
       number = {2},
        pages = {417-430},
          doi = {10.1093/pasj/psab008},
archivePrefix = {arXiv},
       eprint = {2005.04075},
 primaryClass = {astro-ph.GA},
       adsurl = {https://ui.adsabs.harvard.edu/abs/2021PASJ...73..417T}
}

@ARTICLE{Finn2019,
       author = {{Finn}, Molly K. and {Johnson}, Kelsey E. and {Brogan}, Crystal L. and {Wilson}, Christine D. and {Indebetouw}, Remy and {Harris}, William E. and {Kamenetzky}, Julia and {Bemis}, Ashley},
        title = "{New Insights into the Physical Conditions and Internal Structure of a Candidate Proto-globular Cluster}",
      journal = {\apj},
         year = 2019,
        month = apr,
       volume = {874},
       number = {2},
          eid = {120},
        pages = {120},
          doi = {10.3847/1538-4357/ab0d1e},
archivePrefix = {arXiv},
       eprint = {1903.08669},
 primaryClass = {astro-ph.GA},
       adsurl = {https://ui.adsabs.harvard.edu/abs/2019ApJ...874..120F}
}

@ARTICLE{Ueda2012,
       author = {{Ueda}, Junko and {Iono}, Daisuke and {Petitpas}, Glen and {Yun}, Min S. and {Ho}, Paul T.~P. and {Kawabe}, Ryohei and {Mao}, Rui-Qing and {Mart{\'\i}n}, Sergio and {Matsushita}, Satoki and {Peck}, Alison B. and {Tamura}, Yoichi and {Wang}, Junzhi and {Wang}, Zhong and {Wilson}, Christine D. and {Zhang}, Qizhou},
        title = "{Unveiling the Physical Properties and Kinematics of Molecular Gas in the Antennae Galaxies (NGC 4038/9) through High-resolution CO (J = 3-2) Observations}",
      journal = {\apj},
         year = 2012,
        month = jan,
       volume = {745},
       number = {1},
          eid = {65},
        pages = {65},
          doi = {10.1088/0004-637X/745/1/65},
archivePrefix = {arXiv},
       eprint = {1110.2496},
 primaryClass = {astro-ph.GA},
       adsurl = {https://ui.adsabs.harvard.edu/abs/2012ApJ...745...65U}
}

@ARTICLE{Mengel2005,
       author = {{Mengel}, S. and {Lehnert}, M.~D. and {Thatte}, N. and {Genzel}, R.},
        title = "{Star-formation in NGC 4038/4039 from broad and narrow band photometry: cluster destruction?}",
      journal = {\aap},
         year = 2005,
        month = nov,
       volume = {443},
       number = {1},
        pages = {41-60},
          doi = {10.1051/0004-6361:20052908},
archivePrefix = {arXiv},
       eprint = {astro-ph/0505445},
 primaryClass = {astro-ph},
       adsurl = {https://ui.adsabs.harvard.edu/abs/2005A&A...443...41M}
}

@ARTICLE{Whitmore2010,
       author = {{Whitmore}, Bradley C. and {Chandar}, Rupali and {Schweizer}, Fran{\c{c}}ois and {Rothberg}, Barry and {Leitherer}, Claus and {Rieke}, Marcia and {Rieke}, George and {Blair}, W.~P. and {Mengel}, S. and {Alonso-Herrero}, A.},
        title = "{The Antennae Galaxies (NGC 4038/4039) Revisited: Advanced Camera for Surveys and NICMOS Observations of a Prototypical Merger}",
      journal = {\aj},
         year = 2010,
        month = jul,
       volume = {140},
       number = {1},
        pages = {75-109},
          doi = {10.1088/0004-6256/140/1/75},
archivePrefix = {arXiv},
       eprint = {1005.0629},
 primaryClass = {astro-ph.EP},
       adsurl = {https://ui.adsabs.harvard.edu/abs/2010AJ....140...75W}
}
\bibliographystyle{aasjournalv7}



\end{document}